\documentclass[11pt,english]{article}
\usepackage[latin9]{inputenc}
\usepackage{array}
\usepackage{amsmath}
\usepackage{amssymb}
\usepackage{cancel}
\usepackage{graphicx}
\usepackage{setspace}
\makeatletter

\providecommand{\tabularnewline}{\\}

\numberwithin{equation}{section}

\@ifundefined{date}{}{\date{}}
\usepackage{esint}
\usepackage{hyperref}
\usepackage{latexsym}
\usepackage{graphicx}\usepackage{bm}\usepackage{longtable}

\usepackage{xcolor}

\@addtoreset{equation}{section}

\makeatother

\usepackage{babel}
\begin{document}
\title{Low-energy hadronic physics in holographic QCD\textsubscript{3} with
anisotropy}
\maketitle
\begin{center}
Si-wen Li\footnote{Email: siwenli@dlmu.edu.cn}, 
\par\end{center}

\begin{center}
\emph{Department of Physics, School of Science,}\\
\emph{Dalian Maritime University, }\\
\emph{Dalian 116026, China}\\
\par\end{center}

\vspace{12mm}

\begin{abstract}
Using the gauge-gravity duality, we construct the anisotropic D3/D7
approach as a three-dimensional QCD-like theory, then investigate
systematically the hadronic mass spectra, the dragging terms and the
lowest hadronic interactions in the presence of the anisotropy in
holography. Our derivation illustrates the dragging terms in the effective
action are very necessary for an anisotropic theory since they are
the key roles to affect the transport properties of the dual theory.
And the numerical results in addition imply the hadronic system will
become unstable if the anisotropy is much larger than its confinement
energy scale. It agrees with that the confining phase in this approach
becomes unstable if the anisotropy is sufficiently large in our previous
works with this model. Therefore, this work is constructive to understand
the anisotropy in the gauge field theory.
\end{abstract}
\newpage{}

\tableofcontents{}

\section{Introduction}

Quantum chromodynamics (QCD) is the fundamental theory to describe
the strong interaction, however it is usually difficult to solve at
low energy because of its asymptotic freedom, especially in the dense
matter or at finite temperature. Fortunately, the gauge-gravity duality,
which is a ``rosetta stone'' in the development of string theory
\cite{Taubes:1999bv,Maldacena:1997re,Aharony:1999ti,Witten:1998qj},
provides an alternative option to study the dynamics of gauge theories
in the strongly coupled region. On the other hand, the QCD matters
or quark-gluon plasma (QGP) created in the heavy-ion collision experiments
is turned out to be strongly coupled \cite{Shuryak:2003xe,Shuryak:2004cy}
and anisotropic \cite{Florkowski:2010cf,Ryblewski:2010bs,Martinez:2010sc,Martinez:2010sd},
therefore investigating the anisotropic and strongly coupled plasma
or Yang-Mills theory by constructing the type IIB supergravity in
holography becomes naturally attractive \cite{Giataganas:2017koz,Banks:2016fab,Avila:2016mno,Li:2022wwv,Li:2021vve,Hong:2010sb,Argurio:2020her,Li:2025ahp}
since the correspondence between four-dimensional $\mathcal{N}=4$
$SU\left(N_{c}\right)$ super Yang-Mills theory on $N_{c}$ D3-branes
and type IIB super string theory on $\mathrm{AdS_{5}}\times S^{5}$
is the most famous example in the gauge-gravity duality.

One of the remarkable frameworks in the top-down holographic approaches
including the anisotropy is proposed in \cite{Mateos:2011tv} in which
the black D3-brane solution is spatially anisotropic due to the presence
of the axion field or the inhomogeneously dissolved D7-branes in the
bulk. This gravity solution corresponds to the energy-momentum tensor
with anisotropic pressures in the dual theory. In this sense, the
thermodynamics and transport properties of the anisotropic plasma
can be explored in holography by using the gravity solution which
has attracted wide attentions \cite{Rebhan:2011vd,Mateos:2011ix}.
In addition, the presented axion field in \cite{Mateos:2011tv} leads
to an anisotropic theta term $\theta\int F\wedge F$ in the dual theory
which involves the topological property or CP violation in the strong
interaction, so the framework in \cite{Mateos:2011tv} also relates
to various topics about the theta term or instantons in QCD \cite{Schafer:1996wv,Gross:1980br,Tong:2005un}
e.g. the deconfinement phase transition \cite{DElia:2012pvq,DElia:2013uaf,Vicari:2008jw},
$P$ or $CP$ violation in hadronic decays \cite{HFLAV:2022esi,LHCb:2025ray}. 

With all of the above, it is motivated sufficiently to investigate
systematically the low-energy hadronic physics with this anisotropic
model in holography. In this work, we follow the common steps in \cite{Hong:2010sb,Becker:2007,Witten:1998zw}
first in order to construct the frameworks in \cite{Mateos:2011tv}
as a three-dimensional confining Yang-Mills theory with anisotropy.
Then we embed the D7-branes (and $\overline{\mathrm{D7}}$-branes)
as flavors into the anisotropic confining bulk geometry, so the dual
theory becomes a three-dimensional QCD-like theory (QCD\textsubscript{3}),
and the effective actions for the scalar and vector mesons are derived
accordingly. Afterwards, the baryon vertex is introduced as a D5-brane
wrapped on $S^{5}$ as it is discussed in \cite{Witten:1998xy}, so
the fermionic flux created by the $N_{c}$ open strings connecting
to the baryon vertex in the bulk can be identified to be a baryonic
field and its effective action can be obtained by using the T-duality
rules for the worldvolume fermion in string theory \cite{Li:2025ahp,Li:2023wyb,Nakas:2020hyo,Marolf:2003ye,Marolf:2003vf}.
Accordingly, we evaluate numerically the mass spectra of various mesons
and baryons with respect to the anisotropy, and furthermore find noticeably
the gradient mixing or intergradient coupling terms induced by the
anisotropy presented in the dual theory action which describes the
drag effect among the various mesons and baryons in the anisotropic
media. The gradient mixing terms in general are known to be necessary
in various theories with anisotropy e.g. in the anisotropic Ginzburg-Landau
theory for superconductor \cite{PhysRevLett.83.5350,physleta.2016}.
Distinct to the phenomenological theories, all the dragging coefficients
and the mass spectra in our holographic model can be determined automatically
without any additional parameters. Our numerical analysis illustrate
that the bosonic mesons would include imaginary frequency thus becomes
unstable when the anisotropic parameter $a$ is close to the deconfined
energy $M_{KK}$. And in this situation, the interaction terms in
the actions involving fermionic baryon dominate the hadronic interaction.
These behaviors are opposite to the analysis of the D3/D7 model with
isotropic instantons \cite{Li:2025ahp,Li:2024jkd}, therefore the
presence of the anisotropy would induce additional instabilities to
the hadronic system according to the analysis from the gauge-gravity
duality.

The outline of this work is as follows. In Section 2, anisotropic
black brane solution in the type IIB supergravity is reviewed, and
we specify our holographic setup with the anisotropic solution. In
Section 3, we derive the canonical kinetic action with dragging terms
for worldvolume bosons and fermions on the flavor D7-branes as mesons
and baryons, then evaluate numerically the mass spectra and the dragging
coefficients. In Section 4, we derive the interaction terms involving
the holographic mesons and baryons and evaluate numerically the corresponding
coupling constants as functions of $a/M_{KK}$. In Section 5, we give
the summary of this work and our discussion. In the appendixes, we
give the regular functions used in the text, the full formulas of
the D-brane action and all the terms for the actions involving the
fermionic baryons in the articles, which would be useful to this work. 

\section{The holographic setup }

\subsection{The anisotropic solution from type IIB supergravity}

In this section, let us collect the relevant contents of the anisotropic
solution in the type IIB supergravity \cite{Mateos:2011tv}. This
solution describes the dynamics of $N_{c}$ D3-branes with $N_{\mathrm{D7}}$
D7-branes dissolved inhomogeneously in the ten-dimensional bulk spacetime
in the limit of $N_{c}\rightarrow\infty$ while $N_{\mathrm{D7}}/N_{c}$
is fixed. So the backreaction of the $N_{\mathrm{D7}}$ D7-branes
can not be neglected and the D-brane configuration is given in Table
\ref{tab:1}. 
\begin{table}[h]
\begin{centering}
\begin{tabular}{|c|c|c|c|c|c|c|}
\hline 
Black brane background & $t$ & $x$ & $y$ & $z$ & $u$ & $\Omega_{5}$\tabularnewline
\hline 
\hline 
$N_{c}$ D3-branes & - & - & - & - &  & \tabularnewline
\hline 
$N_{\mathrm{D7}}$ D7-branes & - & - & - &  &  & -\tabularnewline
\hline 
\end{tabular}
\par\end{centering}
\caption{\label{tab:1} The configuration of the D-branes in the black brane
background. ``-'' represents the D-brane extends along the direction.}
\end{table}
 The concerned action involves the dynamics of a scalar dilaton $\phi$,
metric $g_{MN}$, Ramond-Ramond zero-form $\chi$ and four-form $C_{4}$
which in string frame is given as,

\begin{equation}
S_{\mathrm{IIB}}=\frac{1}{2\kappa_{10}^{2}}\int d^{10}x\sqrt{-g}\left[e^{-2\phi}\left(\mathcal{R}+4\partial_{M}\phi\partial^{M}\phi\right)-\frac{1}{2}F_{1}^{2}-\frac{1}{4\cdot5!}F_{5}^{2}\right],\label{eq:2.1}
\end{equation}
where $\kappa_{10}$ is the ten-dimensional gravitational coupling
constant given by $2\kappa_{10}^{2}=\left(2\pi\right)^{7}l_{s}^{8}$
and the indices $M,N$ run over the ten-dimensional bulk. The anisotropic
solution can be obtained by using the following ansatz in string frame
as,

\begin{align}
ds^{2} & =\frac{L^{2}}{u^{2}}\left(-\mathcal{F}\mathcal{B}dt^{2}+dx^{2}+dy^{2}+\mathcal{H}dz^{2}+\frac{du^{2}}{\mathcal{F}}\right)+L^{2}\mathcal{Z}d\Omega_{5}^{2},\nonumber \\
F_{1} & =d\chi,\ \chi=az,\ F_{5}=dC_{4}=\frac{4}{L}\left(\Omega_{S^{5}}+\star\Omega_{S^{5}}\right),\nonumber \\
\mathcal{H} & =e^{-\phi},\ \mathcal{Z}=e^{\frac{1}{2}\phi},\label{eq:2.2}
\end{align}
where $L,\Omega_{S^{5}}$ refers to the the radius of the bulk and
the volume form of a five-dimensional sphere $S^{5}$. The presented
parameters are given as follows, 

\begin{equation}
L^{4}=4\pi g_{s}N_{c}l_{s}^{4}=\lambda l_{s}^{4},a=\frac{\lambda n_{\mathrm{D7}}}{4\pi N_{c}},\label{eq:2.3}
\end{equation}
where $g_{s},\lambda$ represent the string coupling and the 't Hooft
coupling constant respectively. Note that the $N_{\mathrm{D7}}$ D7-branes
are distributed along $z$ direction and its distribution density
$n_{\mathrm{D7}}=dN_{\mathrm{D7}}/dz$ is constant according to the
solution for $\chi$, thus the parameter $a$ characterizes the anisotropy
in $z$ direction in this system. As the dissolved $N_{\mathrm{D7}}$
D7-branes do not extend along the holographic direction $u$, thus
there is not new fields arising in the holographic boundary at the
$u=0$. Taking into account the axion field $\chi$ magnetically coupling
to the $N_{\mathrm{D7}}$ D7-branes, it is clear that the solution
(\ref{eq:2.2}) describes the black D3-branes with a horizon at $u=u_{H}$
involving the backreaction of $N_{\mathrm{D7}}$ D7-branes to the
ten-dimensional bulk geometry. 

In the solution (\ref{eq:2.2}), the presented regular functions $\mathcal{F},\mathcal{B},\phi$
depended only on the holographic coordinate $u$ have to be determined
by their equations of motion. However they are non-analytical functions
in general \cite{Avila:2016mno,Li:2022wwv,Mateos:2011tv}. To avoid
the conical singularities in the bulk, we can find the Euclidean version
of the bulk metric near the horizon $u=u_{T}$ as,

\begin{equation}
ds_{E}^{2}\simeq\frac{1}{u_{H}^{2}}\left[\mathcal{F}^{\prime}\left(u_{H}\right)\mathcal{B}\left(u_{H}\right)\left(u-u_{H}\right)\left(dt_{E}\right)^{2}+\frac{du^{2}}{\mathcal{F}^{\prime}\left(u-u_{H}\right)}\right],\ \mathcal{F}^{\prime}=-\frac{d\mathcal{F}}{du},\label{eq:2.4}
\end{equation}
so that the period $\delta t_{E}$ must be $2\pi$. Hence it leads
to the Hawking temperature $T$ as,

\begin{equation}
\delta t_{E}=\frac{4\pi}{\mathcal{F}^{\prime}\left(u_{H}\right)\sqrt{\mathcal{B}\left(u_{H}\right)}}=\frac{1}{T}.
\end{equation}
In the case of the sufficiently large temperature $T\rightarrow\infty$
(or equivalently $\delta t_{E}\rightarrow0$), the functions $\mathcal{F},\mathcal{B},\phi$
become approximately analytical as the series of $a/T$ ($a/T\ll1$)
given in the Appendix A which leads to a relation as,

\begin{equation}
T=\frac{1}{\pi u_{T}}+\frac{u_{T}}{48\pi}\left(5\log2-2\right)a^{2}+\mathcal{O}\left(a^{4}\right).\label{eq:2.6}
\end{equation}
And this high-temperature analysis would be widely used in the rest
of this work. Besides, according to the fluid/gravity duality \cite{Avila:2016mno,Mateos:2011tv},
the dual theory can be described by the energy-momentum tensor $T_{\mu\nu}$
with anisotropic pressure $P_{i}$ as $T_{\mu\nu}=\mathrm{diag}\left\{ E,P_{x},P_{y},P_{z}\right\} $.
Here $P_{x}=P_{y}\neq P_{z}$ are functions of the anisotropic parameter
$a$ which illustrates specifically the anisotropy in the theory.

\subsection{Construction for a confined theory}

It is well-know that the type IIB supergravity theory corresponds
to the $\mathcal{N}=4$ super Yang-Mills theory living on D3-brane
in gauge-gravity duality \cite{Maldacena:1997re,Aharony:1999ti},
however the $\mathcal{N}=4$ super Yang-Mills theory is a conformal
theory instead of a confined theory. Since our concern of the dual
theory is the QCD-like theory as a confined theory, let us demonstrate
the construction for a confined theory without supersymmetry in the
IIB supergravity with anisotropy. The steps to construct a confined
theory in IIB supergravity can also be reviewed in \cite{Witten:1998qj,Becker:2007}.

Specifically, the first step is to compactify the direction $y$ illustrated
in Table \ref{tab:1} of the D3-branes on a circle $S^{1}$ with the
period $\delta y$, so the dual theory becomes effectively three-dimensional
below the Kaluza-Klein scale $M_{KK}=2\pi/\delta y$. The second step
is to impose the periodic and anti-periodic boundary condition on
the gauge boson and the adjoint fermionic fields respectively along
$S^{1}$ on the D3-brane, so it is equivalent to get rid of all massless
fields other than the gauge fields in the low-energy dynamics. Then,
the dual theory becomes three-dimensional pure gauge theory with confinement
below $M_{KK}$. The final step is to identify the bulk geometry which
corresponds to this confining theory. And the answer can be obtained
by interchanging $t$ and $y$ i.e. performing equivalently a double
Wick rotation as $t\rightarrow-iy,y\rightarrow-it$ to the metric
presented in (\ref{eq:2.2}). We further rename the spatial directions
$x,y,z$ as $x,z,y$, in this sense the confined metric can be obtained
as,

\begin{equation}
ds^{2}=\frac{L^{2}}{u^{2}}\left(-dt^{2}+dx^{2}+\mathcal{H}dy^{2}+\mathcal{F}\mathcal{B}dz^{2}+\frac{du^{2}}{\mathcal{F}}\right)+L^{2}\mathcal{Z}d\Omega_{5}^{2}.\label{eq:2.7}
\end{equation}
Note that the axion field becomes $\chi=ay$ while the other fields
$\phi,F_{5}$ presented in (\ref{eq:2.2}) remain. In the confined
solution (\ref{eq:2.7}), $z$ is periodic with the period $\delta z$
as,

\begin{equation}
\delta z=\frac{4\pi}{\mathcal{F}_{1}\left(u_{KK}\right)\sqrt{\mathcal{B}\left(u_{KK}\right)}}=\frac{2\pi}{M_{KK}},
\end{equation}
to avoid the conical singularities on $z-u$ place, and

\begin{equation}
M_{KK}=\frac{2}{u_{KK}}+\frac{u_{KK}}{24}\left(5\log2-2\right)a^{2}+\mathcal{O}\left(a^{4}\right).
\end{equation}

We have used $u_{KK}$ to replace $u_{T}$ presented in (\ref{eq:2.4}).
The solution (\ref{eq:2.7}) is defined only in the region of $0\leq u\leq u_{KK}$
which represents a bubble geometry without a horizon and is anisotropic
on $\left\{ x,y\right\} $ plane. The D-brane configuration for the
bubble solution (\ref{eq:2.7}) is given in Table \ref{tab:2}. 
\begin{table}
\begin{centering}
\begin{tabular}{|c|c|c|c|c|c|c|}
\hline 
Bubble background & $t$ & $x$ & $y$ & $\left(z\right)$ & $u$ & $\Omega_{5}$\tabularnewline
\hline 
\hline 
$N_{c}$ D3-branes & - & - & - & - &  & \tabularnewline
\hline 
$N_{\mathrm{D7}}$ D7-branes & - & - &  & - &  & -\tabularnewline
\hline 
\end{tabular}
\par\end{centering}
\caption{\label{tab:2} The configuration of the D-branes for the bubble solution
(\ref{eq:2.7}).}
\end{table}
 Note that the high temperature limit $\delta t_{E}\rightarrow0$
now becomes $\delta z\rightarrow0$ or, equivalently $M_{KK}\rightarrow\infty$,
$a/M_{KK}\ll1$, so in this limit the regular functions $\mathcal{F},\mathcal{B},\phi$
remain to be approximately analytical as they are given in the Appendix
A, and the dual theory trends to be purely three-dimensional\footnote{We consider the case of $1\ll a\ll M_{KK}$ in this work to avoid
the instability in the gravity background\cite{Giataganas:2017koz,Mateos:2011tv}.}.

On the other hand, to see the color sector of the dual theory under
the confined geometry (\ref{eq:2.7}) is an obviously three-dimensional
Yang-Mills-Chern-Simons theory at zero temperature, it is possible
to introduce a probe D3-brane located at the holographic boundary
at $u\rightarrow0$ \cite{Li:2022wwv,Hong:2010sb,Argurio:2020her}.
Since the low-energy modes in the dual field theory contain the gauge
fields only, the effective action for a probe D3-brane is given by
the DBI (Dirac-Born-Infeld) plus WZ (Wess-Zumino) action as

\begin{equation}
S_{\mathrm{D3}}=-T_{3}\mathrm{Tr}\int d^{3}xdze^{-\phi}\sqrt{-\det\left(g_{ab}+\mathcal{F}_{ab}\right)}+\frac{1}{2}T_{3}\mathrm{Tr}\int\chi\mathcal{F}\wedge\mathcal{F}.\label{eq:2.10}
\end{equation}
The tension of the D3-brane is $T_{3}=\left(2\pi\right)^{-3}l_{s}^{-4}g_{s}^{-1}$,
and the induced metric, the gauge field strength on the D3-brane worldvolume
is denoted by $g_{ab},\mathcal{F}_{ab}=2\pi\alpha^{\prime}F_{ab}$.
Assuming $\mathcal{F}_{ab}$ does not depend on $z$ and does not
have components along $z$, so the quadratic expansion of the action
(\ref{eq:2.10}) at the holographic boundary would be

\begin{align}
S_{\mathrm{D3}} & =-\frac{1}{4}T_{3}\left(2\pi\alpha^{\prime}\right)^{2}\mathrm{Tr}\int_{\mathrm{D3}}d^{3}xdz\sqrt{-g}e^{-\phi}g^{ac}g^{bd}F_{ab}F_{cd}-\mu_{3}\left(2\pi\alpha^{\prime}\right)^{2}\mathrm{Tr}\int_{\mathrm{D3}}d\chi\wedge\omega_{3}\nonumber \\
 & =-\frac{N_{c}}{4\lambda_{3}}\mathrm{Tr}\int_{\mathbb{R}^{2+1}}d^{3}xF_{ab}^{2}-\frac{N_{\mathrm{D7}}}{4\pi}\mathrm{Tr}\int_{\mathbb{R}^{2+1}}\omega_{3},\label{eq:11}
\end{align}
where $\lambda_{3}=\lambda M_{KK}/\left(2\pi\right)$ refers to the
three-dimensional 't Hooft coupling constant and 

\begin{equation}
\omega_{3}=A\wedge dA+\frac{2}{3}A\wedge A\wedge A,
\end{equation}
is the Chern-Simons three-form. We have used $A$ to refer to the
gauge potential and have imposed the boundary value of the functions
$\mathcal{F},\mathcal{B},\phi$ as $\phi_{\mathrm{bdry}}=0,\mathcal{F}_{\mathrm{bdry}}=\mathcal{B}_{\mathrm{bdry}}=1$
on (\ref{eq:11}). Remarkably, the level number of the Chern-Simons
term $N_{\mathrm{D7}}$ is integer automatically in this holographic
setup since it is exactly the number of D7-branes in the gravity side.

\subsection{Embedding the $\mathrm{D7/\overline{D7}}$-branes as flavors}

To include the flavors in the dual theory, let us discuss the embedding
of the D7-branes as flavors in the anisotropic background, as the
most approaches of the D3/D7 system \cite{Casalderrey-Solana:2011dxg}.

We consider $N_{f}$ ($N_{f}\ll N_{c}$) coincident $\mathrm{D7/\overline{D7}}$-branes
as probes embedded to the confined anisotropic background (\ref{eq:2.7}),
which span the $\mathbb{R}^{1,2}$ denoted by $\left\{ t,x,y\right\} $,
the holographic direction denoted by $u$, four of the five directions
in $\Omega_{5}$ \cite{Hong:2010sb,Li:2025ahp,Casalderrey-Solana:2011dxg,Kruczenski:2003uq}
and are perpendicular to the compactified $z$ direction. The relevant
configuration including various D-branes is given in Table \ref{tab:3}.
Note that the directions of $\Omega_{5}$ presented in (\ref{eq:2.7})
has been decomposed as $\Omega_{4}$ and $w$. 
\begin{table}
\begin{centering}
\begin{tabular}{|c|c|c|c|c|c|c|c|}
\hline 
Bubble background & $t$ & $x$ & $y$ & $\left(z\right)$ & $u$ & $\Omega_{4}$ & $w$\tabularnewline
\hline 
\hline 
$N_{c}$ D3-branes & - & - & - & - &  &  & \tabularnewline
\hline 
$N_{\mathrm{D7}}$ D7-branes & - & - &  & - &  & - & -\tabularnewline
\hline 
$N_{f}$ D7-branes & - & - & - &  & - & - & \tabularnewline
\hline 
Baryon vertex D5-branes & - &  &  &  &  & - & -\tabularnewline
\hline 
\end{tabular}
\par\end{centering}
\caption{\label{tab:3} The D-brane configuration for various D-branes with
respect to the bubble geometry.}
\end{table}
 Quarks in this setup are defined as the lowest fermionic excitation
of the open strings connecting the flavor D7- and color D3- branes,
hence they are fundamental representations of the flavor $U\left(N_{f}\right)$
and color $U\left(N_{c}\right)$ group. The leftover direction $w$
is perpendicular to both the $N_{f}$ flavor D7-branes and the $N_{c}$
color D3-branes, so it implies a bare mass for the fundamental quarks
can be introduced by the finite separation between color and flavor
branes at the UV boundary. And the parity in QCD\textsubscript{3}
is therefore broken. In our D3/D7 approach, the excitation of the$w$
direction corresponds to the scalar mesonic field on the D7-brane
worldvolume which couples to the mass operator of the fermions. According
to the gauge-gravity duality, the profile along the transverse direction
of the flavor branes can furthermore correspond to the mesonic operator
$\bar{\psi}\psi$ in the dual theory \cite{Hong:2010sb,Li:2025ahp,Casalderrey-Solana:2011dxg,Kruczenski:2003uq}.

Follow the similar steps given in \cite{Avila:2016mno,Li:2022wwv},
we define a radial coordinate $\rho$ as,

\begin{equation}
\frac{du^{2}}{u^{2}\mathcal{F}}=\frac{\mathcal{Z}}{\rho^{2}}d\rho^{2}.
\end{equation}
In the limit of $\delta z\rightarrow0$ or $a/M_{KK}\ll1$, the relation
of $u$ and $\rho$ is worked out, up to order of $\mathcal{O}\left(a^{2}\right)$,
as,

\begin{align}
u\left(\rho\right)= & u_{0}\left(\rho\right)+a^{2}u_{2}\left(\rho\right),\nonumber \\
u_{0}\left(\rho\right)= & \frac{2L^{2}u_{KK}^{2}\rho}{\sqrt{L^{8}+4u_{KK}^{4}\rho^{4}}},\nonumber \\
u_{2}\left(\rho\right)= & -\frac{L^{2}u_{KK}^{4}\rho}{24\left(L^{8}+4u_{KK}^{4}\rho^{4}\right)^{3/2}}\bigg[4L^{4}u_{KK}^{2}\rho^{2}+2L^{8}\left(\log32-1\right)\nonumber \\
 & -5\left(L^{8}+4u_{KK}^{4}\rho^{4}\right)\log\left(\frac{L^{8}+4u_{KK}^{2}\rho^{2}L^{4}+4u_{KK}^{4}\rho^{4}}{L^{8}+4u_{KK}^{4}\rho^{4}}\right)\bigg].\label{eq:2.14}
\end{align}
To avoid the ambiguity in this relation, we can require $\rho\geq L^{2}/\left(\sqrt{2}u_{KK}\right)$
for $a\geq0$, then the background metric (\ref{eq:2.7}) can be written
as,

\begin{equation}
ds^{2}=\frac{L^{2}}{u^{2}}\left(-dt^{2}+dx^{2}+\mathcal{H}dy^{2}+\mathcal{F}\mathcal{B}dz^{2}\right)+\frac{L^{2}\mathcal{Z}}{\rho^{2}}\left(d\rho^{2}+\rho^{2}d\Omega_{5}^{2}\right),\label{eq:2.15}
\end{equation}
where $u$ is a function of $\rho$ as$u=u\left(\rho\right)$. Afterwards,
let us pick up $\Theta$ as one angular coordinate in $\Omega_{5}$,
define the radial as $\zeta=\rho\cos\Theta,w=\rho\sin\Theta$ or equivalently
$\rho^{2}=w^{2}+\zeta^{2}$, thus the metric (\ref{eq:2.15}) takes
the form

\begin{equation}
ds^{2}=\frac{L^{2}}{u^{2}}\left(-dt^{2}+dx^{2}+\mathcal{H}dy^{2}+\mathcal{F}\mathcal{B}dz^{2}\right)+\frac{L^{2}\mathcal{Z}}{\rho^{2}}\left(d\zeta^{2}+\zeta^{2}d\Omega_{4}^{2}+dw^{2}\right).
\end{equation}
Due to the \ref{tab:3}, the flavor D7-branes extend along $\left\{ t,x,y,u,\Omega_{4}\right\} $,
so the induced metric on the D7-branes would be,

\begin{equation}
ds_{\mathrm{D7}}^{2}=\frac{L^{2}}{u^{2}}\left(-dt^{2}+dx^{2}+\mathcal{H}dy^{2}\right)+\frac{L^{2}}{u^{2}}\mathcal{F}\mathcal{B}z^{\prime2}d\zeta^{2}+\frac{L^{2}\mathcal{Z}}{\rho^{2}}\left[\left(w^{\prime2}+1\right)d\zeta^{2}+\zeta^{2}d\Omega_{4}^{2}\right],\label{eq:2.17}
\end{equation}
where $z=z\left(\zeta\right),w=w\left(\zeta\right)$ and the derivatives
`` $^{\prime}$ '' are with respect to $\zeta$. In this sense,
the embedding functions of the D7-branes can be determined by solving
the equations of motion for $z\left(\zeta\right),w\left(\zeta\right)$
by using the D7-brane action

\begin{equation}
S_{\mathrm{D7}}=-T_{7}\int d^{8}xe^{-\phi}\sqrt{-g_{\mathrm{D7}}},
\end{equation}
which can be reviewed in \cite{Li:2022wwv,Li:2021vve}. As a result,
it is possible to chose a constant solution for $z\left(\zeta\right)$.
Meanwhile, to chose $w\left(\zeta\right)=0$ as the solution is also
possible while it requires that the D7- and $\mathrm{\overline{D7}}$-
branes take opposite orientations located at the antipodal position
of $z$ direction and their worldvolume is connected at $u=u_{KK}$.
Overall, the embedding of the $\mathrm{D7/\overline{D7}}$-branes
can follow the similar discussions as the $\mathrm{D4/}\mathrm{D8/\overline{D8}}$
and $\mathrm{D4/}\mathrm{D6/\overline{D6}}$ approaches \cite{Kruczenski:2003uq,Sakai:2004cn}.
We will use the constant solution for $z\left(\zeta\right)$ and $w\left(\zeta\right)=0$
as the embedding solution of the D7-branes for simplification.

\subsection{The baryon vertex}

\begin{figure}
\begin{centering}
\includegraphics[scale=0.35]{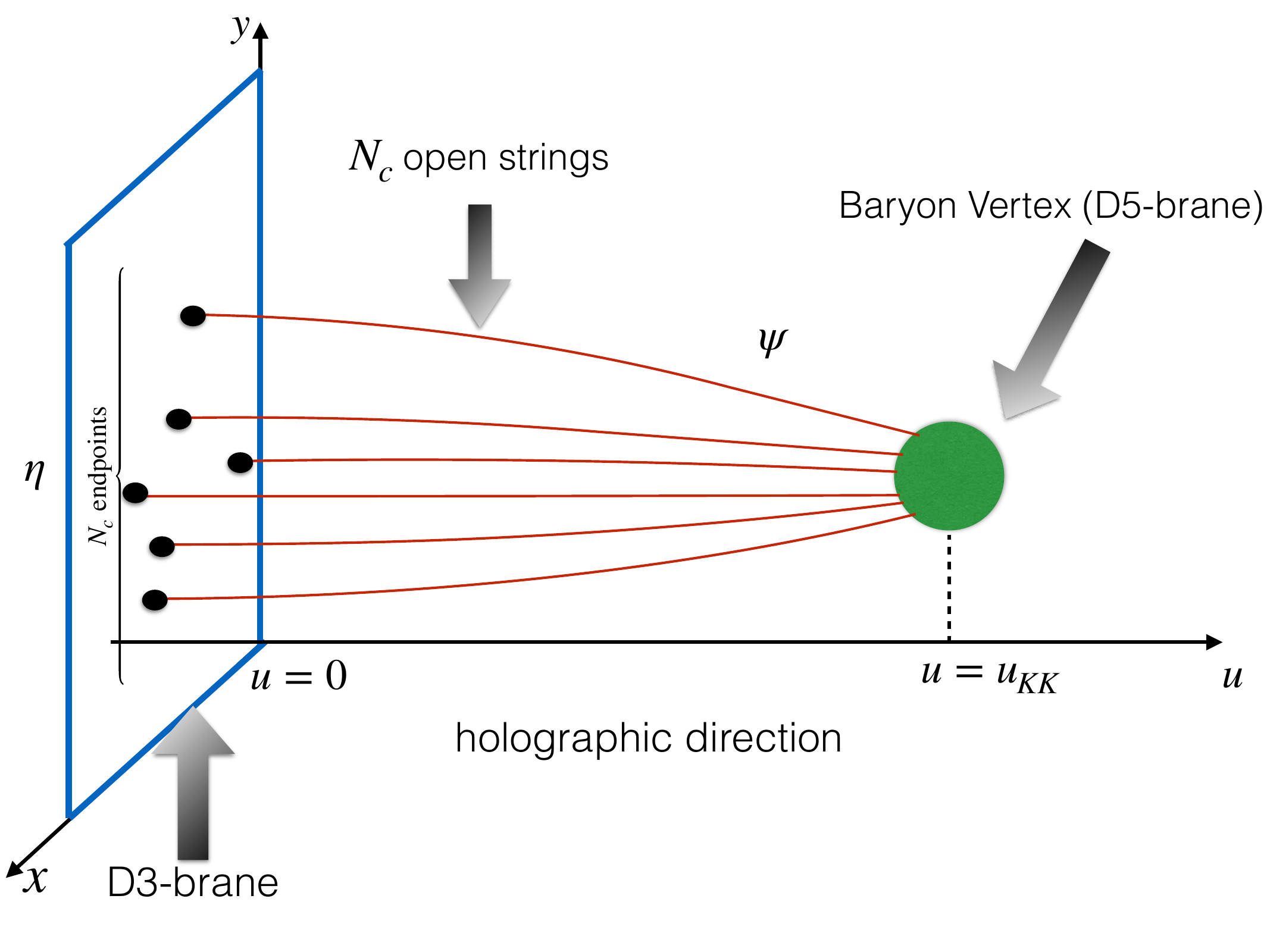}
\par\end{centering}
\caption{\label{fig:1}The configuration of the baryon vertex.}
\end{figure}

Here, we briefly outline the holographic duality of the baryon incorporating
a baryon vertex with respect to our D3/D7 setup. According to the
framework of gauge-gravity duality, a baryon vertex corresponds to
a probe D-brane that wraps on the dimensions specified by the spherical
part of the background geometry, with $N_{c}$ open strings terminating
on it \cite{Witten:1998xy}. Therefore, in the context of type IIB
supergravity on $\mathrm{AdS}_{5}\times S^{5}$, the baryon vertex
takes the form as a D5-brane wrapped on $S^{5}$ and possesses $N_{c}$
open strings that end on it extending to the holographic boundary.
The configuration of the D5-brane is given in Table \ref{tab:3}.
Strictly, we place a probe D3-brane at the holographic boundary, then,
as depicted in Figure \ref{fig:1}, there exist open strings of identical
orientation stretched to the holographic boundary and terminate on
the D5-brane. Owing to the R-R flux contribution $N_{c}\sim\int_{S^{5}}F_{5}$,
the wrapped D5-brane acquires a $U(1)$ charge proportional to $N_{c}$.
On the other hand, each of the $N_{c}$ open strings possesses a unit
$U(1)$ Chan-Paton charge of either $+1$ or $-1$ depending on its
orientation. So their sum can cancel precisely the $U(1)$ charge
residing on the wrapped D5-brane. Thus, by appropriately selecting
the orientations of the D5-brane and the $N_{c}$ open strings, the
baryon current would be conserved at the baryon vertex \cite{Witten:1998xy}. 

In this sense, in the boundary theory, the baryon state $\eta$ is
constructed by the $N_{c}$ endpoints of the $N_{c}$ open strings
as the color singlet. In the bulk, the dual field $\Psi$ is a flux
created by the $N_{c}$ open strings which is gauge-invariant operator
under color group $U\left(N_{c}\right)$. Since in the large $N_{c}$
QCD with the limit of $N_{c}\rightarrow\infty$, baryon is flavored
fermion \cite{Manohar:1998xv,Dashen:1993jt}, hence we consider the
fermionic flux $\Psi$ on the worldvolume of the flavor D7-brane as
the dual field to the baryon $\eta$ living at the boundary. Note
that the fermionic flux $\Psi$ is the adjoint representation of flavor
group $U\left(N_{f}\right)$ which can be identified to a octet state
or a triplet state depended on the choice of the schedule \cite{Li:2025ahp,Li:2023wyb,Li:2024jkd,Li:2025zbj,Li:2024apc}.
Since the action for the worldvolume fermion can be obtained by using
the T-duality in string theory\cite{Marolf:2003ye,Marolf:2003vf},
we will use this holographic setup to investigate the fermionic baryons
in this work.

\section{Analysis of the hadronic spectra}

\subsection{Mesonic boson}

\subsubsection*{Scalar meson}

The mass spectrum of scalar meson can be obtained by deriving the
canonical quadratic form of the scalar action on the flavor D7-branes.
As the transverse mode of the D7-branes, we consider a scalar flux
$W\left(t,x,y,\zeta\right)$ around $w=0$. The quadratic action of
scalar $W$ can be found in Appendix B, which is given as,

\begin{align}
S_{\mathrm{DBI}} & =-\frac{1}{2}T_{7}\left(2\pi\alpha^{\prime}\right)^{2}\int d^{8}\xi e^{-\phi}\sqrt{-g_{\mathrm{D7}}}g^{ab}g_{ww}\partial_{a}W\partial_{b}W\nonumber \\
 & =-\frac{1}{2}\int d^{3}xd\zeta U_{s}\left(\zeta\right)\left[-\left(\partial_{0}W\right)^{2}+\left(\partial_{x}W\right)^{2}+e^{\phi}\left(\partial_{y}W\right)^{2}+P_{s}\left(\zeta\right)W^{\prime2}\right],\label{eq:3.1}
\end{align}
where

\begin{equation}
U_{s}\left(\zeta\right)=\frac{N_{c}L^{4}}{3\pi^{2}}e^{\frac{\phi}{4}}\frac{\zeta^{4}}{u\rho^{7}},P_{s}\left(\zeta\right)=e^{-\frac{\phi}{2}}\frac{\rho^{2}}{u^{2}}.
\end{equation}
Then let us decompose the scalar field $W$ by using a series of complete
functions $h^{\left(n\right)}$ as,

\begin{equation}
W\left(t,x,y,\zeta\right)=\sum_{n=1}^{\infty}\Phi^{\left(n\right)}\left(t,x,y\right)h^{\left(n\right)}\left(\zeta\right),\label{eq:3.3}
\end{equation}
which satisfy the orthogonal normalization condition

\begin{align}
\int d\zeta U_{s}h^{\left(n\right)}h^{\left(m\right)} & =\delta^{mn}M_{KK}^{-1},\nonumber \\
\int d\zeta U_{s}P_{s}\partial_{\zeta}h^{\left(n\right)}\partial_{\zeta}h^{\left(m\right)} & =\Lambda_{n}^{2}M_{KK}\delta^{mn}.
\end{align}
And these normalization conditions lead to the eigen equation for
$h^{\left(n\right)}$ as,

\begin{equation}
-\partial_{\zeta}\left[U_{s}P_{s}\partial_{\zeta}h^{\left(n\right)}\right]=\Lambda_{n}^{2}M_{KK}^{2}U_{s}h^{\left(n\right)},\label{eq:3.5}
\end{equation}
in which the eigen functions $h^{\left(n\right)}$ and its associated
eigen mass can be determined numerically by using the Dirichlet boundary
condition. Afterwards the action (\ref{eq:3.1}) can be integrated
out to be a canonical-like form up to $\mathcal{O}\left(a^{2}\right)$
as

\begin{align}
S_{\mathrm{DBI}}= & S_{\Phi}^{\mathrm{k}}+S_{\Phi}^{\mathrm{c}},\nonumber \\
= & -\frac{1}{2}\int d^{3}xd\zeta U_{s}\sum_{n,m=1}^{\infty}\bigg[-\partial_{0}\Phi^{\left(n\right)}\partial_{0}\Phi^{\left(m\right)}h^{\left(n\right)}h^{\left(m\right)}+\partial_{x}\Phi^{\left(n\right)}\partial_{x}\Phi^{\left(m\right)}h^{\left(n\right)}h^{\left(m\right)}\nonumber \\
 & +e^{\phi}\partial_{y}\Phi^{\left(n\right)}\partial_{y}\Phi^{\left(m\right)}h^{\left(n\right)}h^{\left(m\right)}+P_{s}\Phi^{\left(n\right)}\Phi^{\left(m\right)}\partial_{\zeta}h^{\left(n\right)}\partial_{\zeta}h^{\left(m\right)}\bigg]\label{eq:3.6}
\end{align}
where

\begin{align}
S_{\Phi}^{\mathrm{k}} & =-\frac{1}{2M_{KK}}\int d^{3}x\sum_{n=1}^{\infty}\left[\partial_{\mu}\Phi^{\left(n\right)}\partial^{\mu}\Phi^{\left(n\right)}+\Lambda_{n}^{2}M_{KK}^{2}\Phi^{\left(n\right)}\Phi^{\left(n\right)}\right],\nonumber \\
S_{\Phi}^{\mathrm{c}} & =-\frac{a^{2}}{2M_{KK}^{3}}\int d^{3}x\left[\sum_{n=1}^{\infty}T^{\left(n\right)}\partial_{y}\Phi^{\left(n\right)}\partial_{y}\Phi^{\left(n\right)}+\sum_{n\neq m}^{\infty}S^{\left(m,n\right)}\partial_{y}\Phi^{\left(n\right)}\partial_{y}\Phi^{\left(m\right)}\right],\mu,\nu=0,1,2.\label{eq:3.7}
\end{align}
The presented coefficients are given by the integrals 

\begin{equation}
\int d\zeta U_{s}h^{\left(n\right)}h^{\left(m\right)}\phi_{2}\big|_{a=0}=\begin{cases}
T^{\left(n\right)}M_{KK}^{-3}, & m=n,\\
S^{\left(m,n\right)}M_{KK}^{-3}, & m\neq n.
\end{cases}
\end{equation}
The parity of the scalar can be further determined by taking into
account the Wess-Zumino (WZ) term in the D7-brane action as,

\begin{equation}
S_{\mathrm{WZ}}\sim\int f\wedge f\wedge C_{4},\ C_{4}\sim Wd\Omega_{4},
\end{equation}
where $f$ refers to the gauge field strength on the D7-brane worldvolume
and $d\Omega_{4}$ is the unit volume form of $S^{4}$. Hence $\Phi^{\left(n\right)}$
is a scalar or pseudo-scalar meson when $h^{\left(n\right)}$ is an
even or odd function respectively. We note that while the action $S_{\Phi}^{\mathrm{k}}$
is the canonical kinetic action for the scalars $\Phi^{\left(n\right)}$,
the Lorentz symmetry $SO\left(1,2\right)$ on $\mathbb{R}^{1,2}$
($t,x,y$) is broken down into $SO\left(1,1\right)$ in the presence
of action $S_{\Phi}^{\mathrm{c}}$. This is easy to understand because
the holographic background (\ref{eq:2.7}) is anisotropic along the
$y$-direction. In the effective field theory, the presented action
$S_{\Phi}^{\mathrm{c}}$ is necessary since it describes the drag
effect in the anisotropic media which is also known as gradient mixing
or intergradient coupling \cite{PhysRevLett.83.5350,physleta.2016}
breaking rotational symmetry. The parameters $T^{n}$ describes the
correction of the propagation velocity along the anisotropic direction,
and $S^{\left(m,n\right)}$ describes the dragging with respect to
the other directions. The additional parameters $T^{\left(n\right)},S^{\left(m,n\right)}$
depend on the properties of the anisotropic media and the scalar field
itself which can affect various observables e.g. dispersion relations,
transport coefficients. Different from the phenomenological theories,
in our setup, all the parameters presented in the dual theory (\ref{eq:3.7})
can be determined numerically in holography by using (\ref{eq:3.5}).
\begin{figure}
\begin{centering}
\includegraphics[scale=0.38]{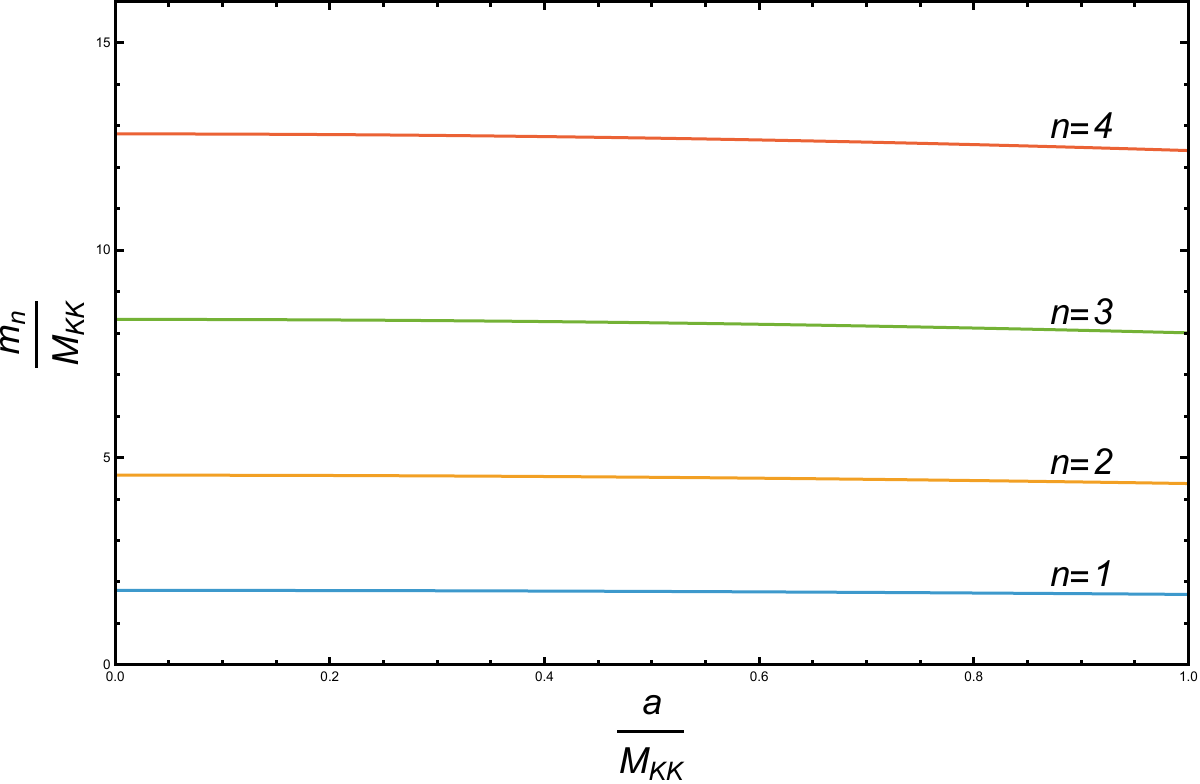}\includegraphics[scale=0.38]{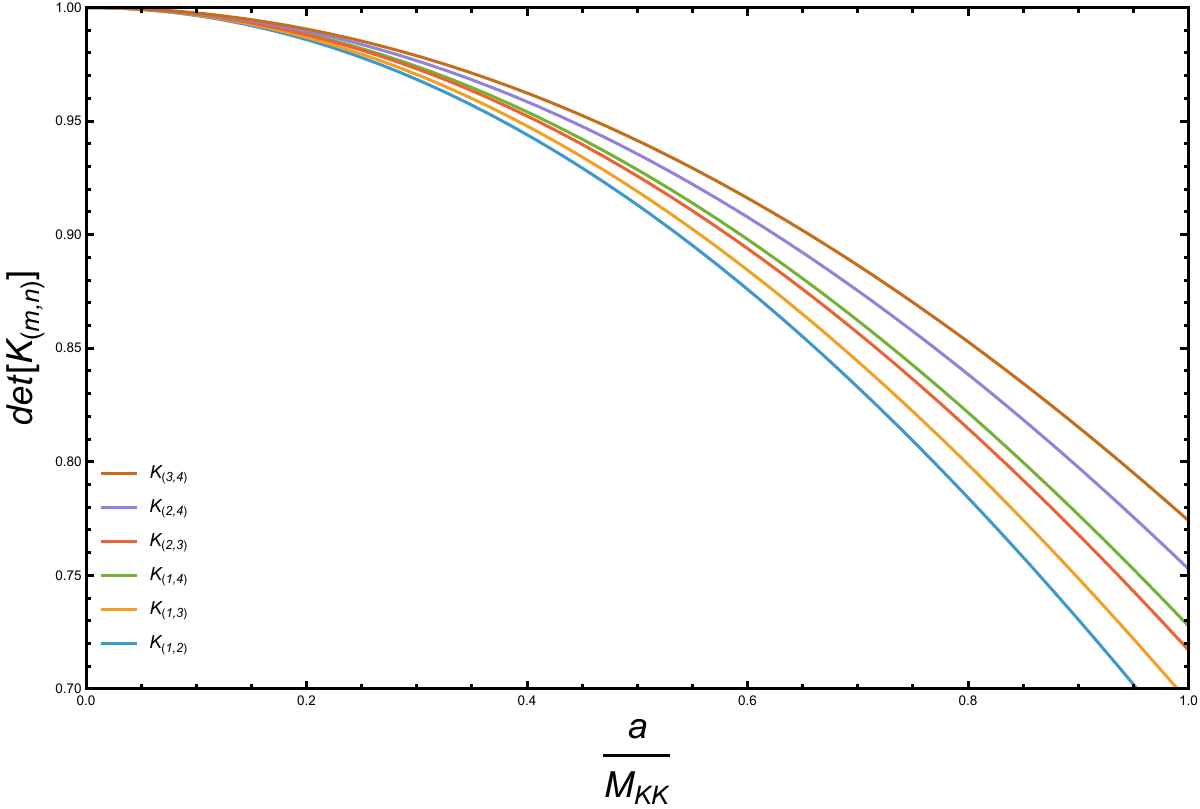}
\par\end{centering}
\caption{\label{fig:2}The mass spectrum $m_{n}=\Lambda_{n}M_{KK}$ of the
scalar meson and $\det K_{\left(m,n\right)}$ as functions of $a/M_{KK}$
for the scalar mesons.}

\end{figure}
 
\begin{table}
\begin{centering}
\begin{tabular}{|>{\centering}p{1.5cm}|>{\centering}p{1.5cm}|>{\centering}p{1.5cm}|>{\centering}p{1.5cm}|>{\centering}p{1.5cm}|>{\centering}p{1.5cm}|>{\centering}p{1.5cm}|}
\hline 
$n$ & $1$ & $2$ & $3$ & $4$ & $5$ & $6$\tabularnewline
\hline 
\hline 
$T^{\left(n\right)}$ & $-0.19$ & $-0.16$ & $-0.14$ & $-0.1$ & $-0.05$ & $-0.04$\tabularnewline
\hline 
\end{tabular}
\par\end{centering}
\begin{centering}
\begin{tabular}{|>{\centering}p{1.5cm}|>{\centering}p{1.5cm}|>{\centering}p{1.5cm}|>{\centering}p{1.5cm}|>{\centering}p{1.5cm}|>{\centering}p{1.5cm}|>{\centering}p{1.5cm}|}
\hline 
$\left(m,n\right)$ & $(1,2)$ & $(1,3)$ & $(1,4)$ & $(2,3)$ & $(2,4)$ & $(3,4)$\tabularnewline
\hline 
\hline 
$S^{\left(m,n\right)}$ & $-0.14$ & $-0.1$ & $0.07$ & $-0.13$ & $0.09$ & $0.1$\tabularnewline
\hline 
\end{tabular}
\par\end{centering}
\caption{\label{tab:4}The lowest dragging coefficients $T^{\left(n\right)},S^{\left(m,n\right)}$
for $m,n\protect\leq6$.}

\end{table}

The mass spectrum and dragging coefficients for the scalar mesons
are given numerically in Figure \ref{fig:2} and Table \ref{tab:4}.
They reveal that the mass of scalar meson is suppressed slightly when
$a/M_{KK}$ increases. This behavior seemingly implies the mass spectrum
trends to be vanished or unstable if $a/M_{KK}$ becomes sufficiently
large. And it would be clear when we focus on the propagation according
to the action. For the anisotropic part, the action presented in (\ref{eq:3.7})
can be written in a more compact form as,

\begin{align}
S_{\mathrm{DBI}} & \sim-\frac{1}{2}\sum_{m,n}\partial_{y}\hat{\Phi}^{T}K_{\left(n,m\right)}\partial_{y}\hat{\Phi},
\end{align}
where

\begin{equation}
\hat{\Phi}=\left(\begin{array}{c}
\Phi^{\left(n\right)}\\
\Phi^{\left(m\right)}
\end{array}\right),\ K_{\left(n,m\right)}=\left(\begin{array}{cc}
1+\frac{a^{2}}{M_{KK}^{3}}T^{\left(n\right)} & \frac{a^{2}}{2M_{KK}^{3}}S^{\left(m,n\right)}\\
\frac{a^{2}}{2M_{KK}^{3}}S^{\left(m,n\right)} & 1+\frac{a^{2}}{M_{KK}^{3}}T^{\left(m\right)}
\end{array}\right).
\end{equation}
Since the matrix $K_{\left(n,m\right)}$ refers to the mixing of the
kinetic terms of the two distinct scalars $\Phi^{\left(n\right)},\Phi^{\left(m\right)}$,
its positive definite would avoid the unstable mesonic modes. So the
conditions for the positive definite of $K_{\left(n,m\right)}$ are

\begin{align}
 & 1+\frac{a^{2}}{M_{KK}^{3}}T^{\left(n\right)}>0,1+\frac{a^{2}}{M_{KK}^{3}}T^{\left(m\right)}>0,\nonumber \\
 & \left[1+\frac{a^{2}}{M_{KK}^{3}}T^{\left(n\right)}\right]\left[1+\frac{a^{2}}{M_{KK}^{3}}T^{\left(m\right)}\right]>\frac{a^{4}}{4M_{KK}^{6}}\left[S^{\left(m,n\right)}\right]^{2}.\label{eq:3.12}
\end{align}
The first line of (\ref{eq:3.12}) can be satisfied definitely if
we consider the case of $a/M_{KK}\ll1$. However, according to the
numerical evaluation given in Figure \ref{fig:2} and Table \ref{tab:4},
the matrix $K_{\left(n,m\right)}$ trends to become non-positive definite
at a critical value of $\frac{a}{M_{KK}}$, thus the holographic system
may include unstable mesonic scalar modes if $\frac{a}{M_{KK}}$ is
large enough.

\subsubsection*{Vector meson}

Let us turn to consider the vector meson. The action for the vector
meson can be obtained by using the DBI action (See the Appendix B)
of the flavor D7-branes involving the gauge field as,

\begin{align}
S_{\mathrm{D7}} & =-T_{\mathrm{D7}}\int d^{8}xe^{-\phi}\sqrt{-\det\left(g_{ab}+2\pi\alpha^{\prime}f_{ab}\right)},\nonumber \\
 & \simeq-T_{\mathrm{D7}}\int_{\mathrm{D7}}d^{3}xd\zeta d\Omega_{4}e^{-\phi}\sqrt{-g_{\mathrm{D7}}}\left[1+\frac{1}{4}\left(2\pi\alpha^{\prime}\right)^{2}g^{ac}g^{bd}f_{ab}f_{cd}+\mathcal{O}\left(\alpha^{\prime}{}^{4}\right)\right],
\end{align}
where $f_{ab}=\partial_{a}A_{b}-\partial_{b}A_{a}-i\left[A_{a},A_{b}\right]$
is the gauge field strength and $A_{a}$ is the gauge field on the
D7-branes. So the quadratic kinetic term of the gauge field strength
reads

\begin{equation}
S_{\mathrm{D7}}=-\frac{1}{4}\left(2\pi\alpha^{\prime}\right)^{2}T_{\mathrm{D7}}\int_{\mathrm{D7}}d^{3}xd\zeta d\Omega_{4}e^{-\phi}\sqrt{-g}g^{ac}g^{bd}f_{ab}f_{cd}.\label{eq:3.14}
\end{equation}
Follow the same steps as we have discussed in the case of the scalar
meson, we assume the gauge potential is $A_{a}=\left\{ A_{\mu}\left(x,\zeta\right),A_{4}\left(x,\zeta\right)\right\} $
and can be decomposed by using a series of complete functions $b^{\left(n\right)}\left(\zeta\right),d^{\left(n\right)}\left(\zeta\right)$
as,
\begin{equation}
A_{\mu}\left(x,\zeta\right)=\sum_{n=1}B_{\mu}^{\left(n\right)}\left(x\right)b^{\left(n\right)}\left(\zeta\right),A_{4}\left(x,\zeta\right)=\sum_{n=1}\varphi^{\left(n\right)}\left(x\right)d^{\left(n\right)}\left(\zeta\right).\label{eq:3.15}
\end{equation}
Then impose (\ref{eq:3.15}) into (\ref{eq:3.14}), the quadratic
action becomes

\begin{align}
S_{\mathrm{D7}}= & -\frac{1}{2}\int d^{3}xd\zeta U_{v}\sum_{n,m=1}^{\infty}\bigg\{-\left[\partial_{t}B_{x}^{\left(n\right)}-\partial_{x}B_{t}^{\left(n\right)}\right]\left[\partial_{t}B_{x}^{\left(m\right)}-\partial_{x}B_{t}^{\left(m\right)}\right]b^{\left(n\right)}b^{\left(m\right)}\nonumber \\
 & -\left[\partial_{t}B_{y}^{\left(n\right)}-\partial_{y}B_{t}^{\left(n\right)}\right]\left[\partial_{t}B_{y}^{\left(m\right)}-\partial_{y}B_{t}^{\left(m\right)}\right]b^{\left(n\right)}b^{\left(m\right)}e^{\phi}\nonumber \\
 & +\left[\partial_{x}B_{y}^{\left(n\right)}-\partial_{y}B_{x}^{\left(n\right)}\right]\left[\partial_{x}B_{y}^{\left(m\right)}-\partial_{y}B_{x}^{\left(m\right)}\right]b^{\left(n\right)}b^{\left(m\right)}e^{\phi}\nonumber \\
 & -\left[\partial_{t}\varphi^{\left(n\right)}d^{\left(n\right)}-B_{t}^{\left(n\right)}\partial_{\zeta}b^{\left(n\right)}\right]\left[\partial_{t}\varphi^{\left(m\right)}d^{\left(m\right)}-B_{t}^{\left(m\right)}\partial_{\zeta}b^{\left(m\right)}\right]P_{s}\nonumber \\
 & +\left[\partial_{x}\varphi^{\left(n\right)}d^{\left(n\right)}-B_{x}^{\left(n\right)}\partial_{\zeta}b^{\left(n\right)}\right]\left[\partial_{x}\varphi^{\left(m\right)}d^{\left(m\right)}-B_{x}^{\left(m\right)}\partial_{\zeta}b^{\left(m\right)}\right]P_{s}\nonumber \\
 & +\left[\partial_{y}\varphi^{\left(n\right)}d^{\left(n\right)}-B_{y}^{\left(n\right)}\partial_{\zeta}b^{\left(n\right)}\right]\left[\partial_{y}\varphi^{\left(m\right)}d^{\left(m\right)}-B^{\left(m\right)}\partial_{\zeta}b^{\left(m\right)}\right]P_{s}e^{\phi}\bigg\},\label{eq:3.16}
\end{align}
where

\begin{equation}
U_{v}\left(\zeta\right)=\frac{N_{c}}{6\pi^{2}}\frac{u\zeta^{4}}{\rho^{5}}e^{-\frac{\phi}{4}},
\end{equation}
Our goal is to obtain the canonical form for the the vector meson
action, so let us further require the orthogonal normalization conditions
for the functions $b^{\left(n\right)}\left(\zeta\right),d^{\left(n\right)}\left(\zeta\right)$
as,

\begin{align}
\int d\zeta U_{v}b^{\left(n\right)}b^{\left(m\right)} & =\delta^{nm}M_{KK}^{-1},\nonumber \\
\int d\zeta U_{v}P_{s}d^{\left(n\right)}d^{\left(m\right)} & =\delta^{nm}M_{KK}^{-1},\nonumber \\
d^{\left(n\right)}=m_{n}^{-1}\partial_{\zeta}b^{\left(n\right)} & ,m_{n}=\Lambda_{n}M_{KK}
\end{align}
which leads to the eigen equation for $b^{\left(n\right)}$ as,

\begin{equation}
-\partial_{\zeta}\left[U_{v}P_{s}\partial_{\zeta}b^{\left(m\right)}\right]=m_{n}^{2}U_{v}b^{\left(m\right)}.\label{eq:3.19}
\end{equation}
Thus the mass spectrum of the vector mesons can be obtained by solving
numerically (\ref{eq:3.19}). Notice that, there is an additional
function $d^{\left(0\right)}=U_{v}^{-1}P_{s}^{-1}$ orthogonal to
all the $d^{\left(n\right)}$'s for $n\geq1$, therefore, the sum
presented in (\ref{eq:3.15}) with respected to $A_{4}$ should include
the terms with $n=0$. On the other hand, since the action (\ref{eq:3.14})
is invariant under the gauge transformation $A_{a}\rightarrow A_{a}-\partial_{a}\chi$,
it is necessary to impose a gauge transformation $B_{\mu}\rightarrow B_{\mu}+m_{n}^{-1}\partial_{\mu}\varphi$
on (\ref{eq:3.16}). Altogether, the action (\ref{eq:3.16}) can be
written up to $\mathcal{O}\left(a^{2}\right)$ as $S_{\mathrm{D7}}=S_{V}^{\mathrm{k}}+S_{V}^{\mathrm{c}},$
where

\begin{align}
S_{V}^{\mathrm{k}}= & -\frac{1}{M_{KK}}\int d^{3}x\sum_{n=1}^{\infty}\left[\frac{1}{4}F_{\mu\nu}^{\left(n\right)}F^{\left(n\right)\mu\nu}+\frac{1}{2}m_{n}^{2}B_{\mu}^{\left(n\right)}B^{\left(n\right)\mu}\right]\nonumber \\
 & -\frac{1}{2M_{KK}}\int d^{3}x\partial_{\mu}\pi\partial^{\mu}\pi,\ \ \varphi^{\left(0\right)}\equiv\pi,\nonumber \\
S_{V}^{\mathrm{c}}= & -\frac{a^{2}}{2M_{KK}^{3}}\int d^{3}x\sum_{n=1}^{\infty}\left[F_{\mu y}^{\left(n\right)}F^{\left(n\right)\mu y}R_{T}^{\left(n\right)}+M_{KK}^{2}B_{y}^{\left(n\right)}B_{y}^{\left(n\right)}Q_{T}^{\left(n\right)}\right]\nonumber \\
 & -\frac{a^{2}}{2M_{KK}^{3}}\int d^{3}x\sum_{n\neq m}^{\infty}\left[F_{\mu y}^{\left(n\right)}F^{\left(m\right)\mu y}R_{S}^{\left(n,m\right)}+M_{KK}^{2}B_{y}^{\left(n\right)}B_{y}^{\left(m\right)}Q_{S}^{\left(m,n\right)}\right]\nonumber \\
 & -\frac{a^{2}}{2M_{KK}^{3}}\int d^{3}x\partial_{y}\pi\partial_{y}\pi Q_{T}^{\left(0\right)}.\label{eq:3.20}
\end{align}
and the associated coefficients are given by

\begin{align}
M_{KK}^{3}\int d\zeta U_{v}b^{\left(n\right)}b^{\left(m\right)}\phi_{2}\big|_{a=0} & =\begin{cases}
R_{T}^{\left(n\right)}, & m=n,\\
R_{S}^{\left(n,m\right)}, & m\neq n,
\end{cases}\nonumber \\
M_{KK}\int d\zeta U_{v}P_{v}d^{\left(n\right)}d^{\left(m\right)}\phi_{2}\big|_{a=0} & =\begin{cases}
Q_{T}^{\left(n\right)}, & m=n,\\
Q_{S}^{\left(n,m\right)}, & m\neq n.
\end{cases}
\end{align}
While the action $S_{V}^{\mathrm{k}}$ is the canonical kinetic term
of the vector bosons, the action $S_{V}^{\mathrm{c}}$ breaks down
the Lorentz symmetry $SO\left(1,2\right)$ describing the dragging
of the various vector bosons in the anisotropic system. For the isotropic
case i.e. $a=0$, $\varphi^{\left(0\right)}\equiv\pi$ is the Nambu-Goldstone
boson, and it remains to be massless although the Lorentz symmetry
$SO\left(1,2\right)$ breaks down into $SO\left(1,1\right)$ in the
presence of $S_{V}^{\mathrm{c}}$. The broken Lorentz symmetry also
induces additional mass to the longitudinal mode $B_{y}$ of the gauge
field $B_{\mu}$, and this can be specified by rewriting $B_{\mu}^{\left(n\right)}=\left\{ B_{\alpha}^{\left(n\right)},\hat{\varphi}^{\left(n\right)}\right\} ,\alpha=0,1$,
where $B_{y}^{\left(n\right)}=\hat{\varphi}^{\left(n\right)}$ is
another scalar, then the action $S_{\mathrm{D7}}=S_{V}^{\mathrm{k}}+a^{2}S_{V}^{\mathrm{c}}$
can be written in term of $\hat{\varphi}^{\left(n\right)}$ as, 
\begin{table}
\begin{centering}
\begin{tabular}{|>{\centering}p{1.5cm}|>{\centering}p{1.5cm}|>{\centering}p{1.5cm}|>{\centering}p{1.5cm}|>{\centering}p{1.5cm}|>{\centering}p{1.5cm}|>{\centering}p{1.5cm}|}
\hline 
$n$ & $1$ & $2$ & $3$ & $4$ & $5$ & $6$\tabularnewline
\hline 
\hline 
$R_{T}^{\left(n\right)}$ & $-0.22$ & $-0.18$ & $-0.15$ & $-0.1$ & $-0.06$ & $-0.04$\tabularnewline
\hline 
$Q_{T}^{\left(n\right)}$ & $-0.13$ & $-0.08$ & $-0.06$ & $-0.05$ & $-0.03$ & $-0.02$\tabularnewline
\hline 
\end{tabular}
\par\end{centering}
\begin{centering}
\begin{tabular}{|>{\centering}p{1.5cm}|>{\centering}p{1.5cm}|>{\centering}p{1.5cm}|>{\centering}p{1.5cm}|>{\centering}p{1.5cm}|>{\centering}p{1.5cm}|>{\centering}p{1.5cm}|}
\hline 
$\left(n,m\right)$ & $\left(1,2\right)$ & $\left(1,3\right)$ & $\left(1,4\right)$ & $\left(2,3\right)$ & $\left(2,4\right)$ & $\left(3,4\right)$\tabularnewline
\hline 
\hline 
$R_{S}^{\left(n,m\right)}$ & $0.16$ & $-0.11$ & $0.08$ & $0.14$ & $-0.1$ & $0.11$\tabularnewline
\hline 
$Q_{S}^{\left(n,m\right)}$ & $0.04$ & $-0.0003$ & $-0.001$ & $0.04$ & $-0.009$ & $0.03$\tabularnewline
\hline 
\end{tabular}
\par\end{centering}
\caption{\label{tab:5}The lowest dragging coefficients $R_{T}^{\left(n\right)},Q_{T}^{\left(n\right)},R_{S}^{\left(m,n\right)},Q_{S}^{\left(m,n\right)}$
for $1\protect\leq m,n\protect\leq6$.}

\end{table}
 
\begin{figure}
\begin{centering}
\includegraphics[scale=0.4]{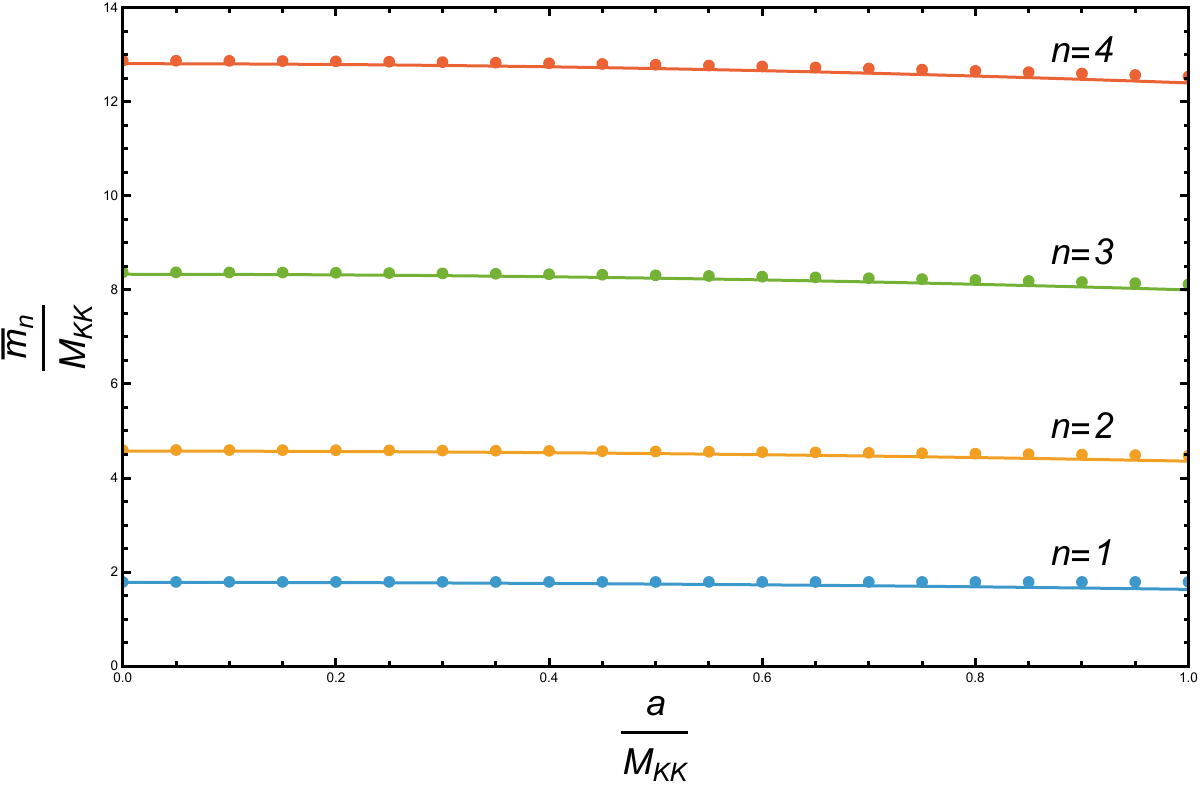}\includegraphics[scale=0.38]{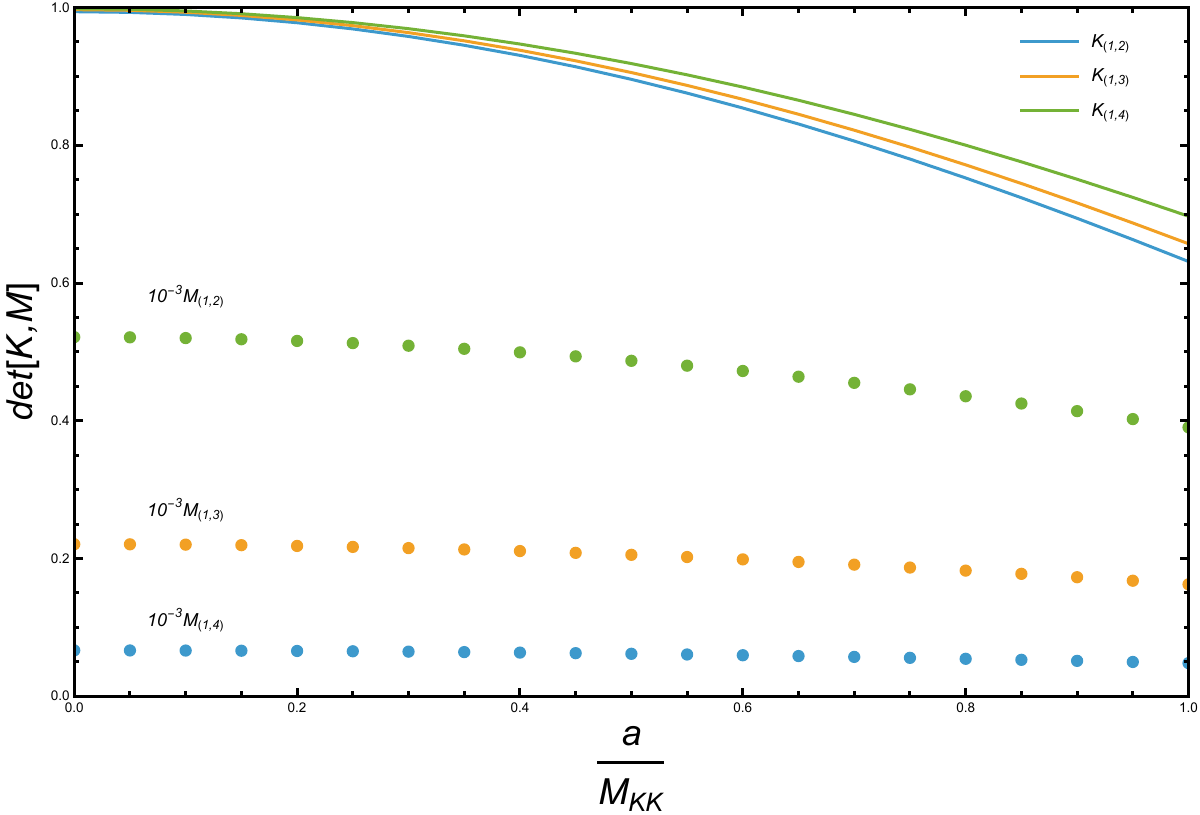}
\par\end{centering}
\caption{\textbf{\label{fig:3}Left:} The mass spectrum $m_{n}$ (dots) and
$\bar{m}_{n}=\sqrt{m_{n}^{2}+a^{2}Q_{T}^{\left(n\right)}}$ (solid
lines) of the vector mesons $B_{\alpha}^{\left(n\right)}$ and scalar
mesons $\hat{\varphi}^{\left(n\right)}$ as functions of $a/M_{KK}$.
\textbf{Right:} The determinant\textcompwordmark{} of matrices $K_{\left(m,n\right)},M_{\left(m,n\right)}$
as functions of $a/M_{KK}$ for the vector mesons. }

\end{figure}

\begin{align}
S_{\mathrm{D7}}= & -\frac{1}{M_{KK}}\int d^{3}x\sum_{n=1}^{\infty}\left[\frac{1}{4}F_{\alpha\beta}^{\left(n\right)}F^{\left(n\right)\alpha\beta}+\frac{1}{2}m_{n}^{2}B_{\alpha}^{\left(n\right)}B^{\left(n\right)\alpha}\right]\nonumber \\
 & -\frac{1}{2M_{KK}}\int d^{3}x\sum_{n=1}^{\infty}\left\{ \left[1+\frac{a^{2}}{M_{KK}^{2}}R_{T}^{\left(n\right)}\right]\partial_{\alpha}\hat{\varphi}^{\left(n\right)}\partial^{\alpha}\hat{\varphi}^{\left(n\right)}+\left[m_{n}^{2}+a^{2}Q_{T}^{\left(n\right)}\right]\hat{\varphi}^{\left(n\right)}\hat{\varphi}^{\left(n\right)}\right\} \nonumber \\
 & -\frac{a^{2}}{2M_{KK}^{3}}\int d^{3}x\sum_{n\neq m}^{\infty}\left[R_{S}^{\left(n,m\right)}\partial_{\alpha}\hat{\varphi}^{\left(n\right)}\partial^{\alpha}\hat{\varphi}^{\left(m\right)}+M_{KK}^{2}Q_{S}^{\left(m,n\right)}\hat{\varphi}^{\left(n\right)}\hat{\varphi}^{\left(m\right)}+...\right].\label{eq:3.22}
\end{align}
Since the coefficients $Q_{T}^{\left(n\right)}$ are corrections to
the mass term, the corrected effective mass of scalar $\hat{\varphi}^{\left(n\right)}$
should be defined as $\bar{m}_{n}=\sqrt{m_{n}^{2}+a^{2}Q_{T}^{\left(n\right)}}$
and the lowest mass spectrum $m_{n}$, coefficients $R_{T}^{\left(n\right)},Q_{T}^{\left(n\right)},R_{S}^{\left(m,n\right)},Q_{S}^{\left(m,n\right)}$
are given numerically in Figure \ref{fig:3} and Table \ref{tab:5}.
As the case of the scalar meson, the mass spectrum of vector meson
is suppressed slightly when $a/M_{KK}$ increases which implies that
the system is unstable if $a/M_{KK}$ is sufficiently large. Moreover,
the action presented in (\ref{eq:3.22}) can be further written in
terms of

\begin{equation}
S_{\mathrm{D7}}\sim\left(\partial_{\alpha}\hat{\varphi}^{\left(n\right)},\partial_{\alpha}\hat{\varphi}^{\left(m\right)}\right)K_{\left(m,n\right)}\left(\begin{array}{c}
\partial_{\alpha}\hat{\varphi}^{\left(n\right)}\\
\partial_{\alpha}\hat{\varphi}^{\left(m\right)}
\end{array}\right)+\left(\hat{\varphi}^{\left(n\right)},\hat{\varphi}^{\left(m\right)}\right)M_{\left(m,n\right)}\left(\begin{array}{c}
\hat{\varphi}^{\left(n\right)}\\
\hat{\varphi}^{\left(m\right)}
\end{array}\right),
\end{equation}
where the kinetic and mass matrices are given as

\begin{equation}
K_{\left(m,n\right)}=\left(\begin{array}{cc}
1+\frac{a^{2}}{M_{KK}^{2}}R_{T}^{\left(n\right)} & \frac{1}{2}R_{S}^{\left(n,m\right)}\\
\frac{1}{2}R_{S}^{\left(n,m\right)} & 1+\frac{a^{2}}{M_{KK}^{2}}R_{T}^{\left(m\right)}
\end{array}\right),\ M_{\left(m,n\right)}=\left(\begin{array}{cc}
m_{n}^{2}+a^{2}Q_{T}^{\left(n\right)} & \frac{1}{2}M_{KK}^{2}Q_{S}^{\left(m,n\right)}\\
\frac{1}{2}M_{KK}^{2}Q_{S}^{\left(m,n\right)} & m_{n}^{2}+a^{2}Q_{T}^{\left(m\right)}
\end{array}\right).
\end{equation}
As before, the system described by the action (\ref{eq:3.22}) would
include the unstable mesonic modes if one of the matrices $K_{\left(m,n\right)},M_{\left(m,n\right)}$
is not positive define. Positive define requires that all the diagonal
elements and the determinants of matrices $K_{\left(m,n\right)},M_{\left(m,n\right)}$
must be positive. Since all the coefficients of $R_{T}^{\left(n\right)}$
($n\leq6$) are negative according to our numerical evaluation in
Table \ref{tab:5}, there are reasons to believe that, at least, the
matrix $K_{\left(m,n\right)}$ will not be positive define if $a/M_{KK}$
is sufficiently large. The numerical evaluation of the determinants
of matrices $K_{\left(m,n\right)},M_{\left(m,n\right)}$ also confirm
this trend. Therefore, all the analysis of the bosonic meson imply
that this holographic system may include unstable modes or become
unstable when $a/M_{KK}$ is sufficiently large.

\subsubsection*{Gauge of $A_{4}=0$}

Since the action (\ref{eq:3.14}) is gauge invariant under the gauge
transformation, an equivalent choice to (\ref{eq:3.15}) can be obtained
by imposing a gauge transformation as,

\begin{equation}
A_{\mu}^{g}=A_{\mu}-\partial_{\mu}\Lambda,A_{4}^{g}=A_{4}-\partial_{4}\Lambda=0,\label{eq:3.25}
\end{equation}
where we have used $A^{g}$ to denote the gauge field under the gauge
condition (\ref{eq:3.25}) and

\begin{align}
\Lambda & =\pi\int d\zeta d^{\left(0\right)}+\sum_{n=1}\varphi^{\left(n\right)}\int d\zeta d^{\left(n\right)}\nonumber \\
 & =\pi b^{\left(0\right)}+\sum_{n=1}m_{n}^{-1}\varphi^{\left(n\right)}b^{\left(n\right)}.
\end{align}
Note that $\int d\zeta d^{\left(0\right)}\equiv b^{\left(0\right)}$.
In this case, we find

\begin{align}
A_{\mu}^{g} & =\sum_{n=1}\left[B_{\mu}^{\left(n\right)}-m_{n}^{-1}\partial_{\mu}\varphi^{\left(n\right)}\right]b^{\left(n\right)}-\partial_{\mu}\pi b^{\left(0\right)}\nonumber \\
 & \equiv\sum_{n=1}B_{\mu}^{g\left(n\right)}b^{\left(n\right)}-\partial_{\mu}\pi b^{\left(0\right)},\label{eq:3.27}
\end{align}
where $B_{\mu}^{g\left(n\right)}\equiv B_{\mu}^{\left(n\right)}-m_{n}^{-1}\partial_{\mu}\varphi^{\left(n\right)}$.
One may check that the choice of (\ref{eq:3.27}) with $A_{4}=0$
induces to (\ref{eq:3.20}), therefore using $A_{4}=0$ gauge would
be alternative to (\ref{eq:3.15}) when it is essential.

\subsection{Baryonic fermion}

In this section, let us investigate the spectrum of the baryonic fermion
in our holographic model. Consider the fermionic flux $\Psi$ on the
flavor D7-branes created by the $N_{c}$ open strings which are connecting
to the baryon vertex as it is illustrated in Figure \ref{fig:1}.
According to the discussion in Section 2.4, $\Psi$ is the baryonic
field in the bulk. Thus the D7-brane action for the worldvolume $\Psi$
is given according to Appendix B with respect to $p=7$ as,

\begin{align}
S_{\mathrm{D7}}^{f}= & \frac{T_{7}N_{c}}{2}\int d^{8}xe^{-\phi}\sqrt{-g}\bar{\Psi}\bigg(\Gamma^{a}\nabla_{a}-\frac{1}{8}e^{\phi}F_{N}\Gamma^{a}\Gamma^{N}\Gamma_{a}\bar{\gamma}\nonumber \\
 & -\frac{e^{\phi}}{2\cdot8\cdot5!}\Gamma^{a}F_{KLMNP}\Gamma^{KLMNP}\Gamma_{a}\bar{\gamma}-\frac{1}{2}\Gamma^{M}\partial_{M}\phi-\frac{1}{2}e^{\phi}F_{M}\Gamma^{M}\bar{\gamma}\bigg)\Psi,\label{eq:3.28}
\end{align}
where the index $a$ runs over the flavor D7-branes and capital letters
$M,N...$ run over the ten-dimensional spacetime. Then the Dirac operator
is computed with respect to geometry on the D7-brane (\ref{eq:2.17})
as ($x^{4}\equiv\zeta$),

\begin{equation}
\Gamma^{a}\nabla_{a}=\frac{u}{L}\gamma^{0}\partial_{0}+\frac{u}{L}\gamma^{1}\partial_{1}+e^{\frac{\phi}{2}}\frac{u}{L}\gamma^{2}\partial_{2}+e^{-\frac{\phi}{4}}\frac{\zeta}{L}\gamma^{4}\partial_{\zeta}+e^{-\frac{\phi}{4}}\frac{\zeta}{uL}\left(\frac{1}{4}u\phi^{\prime}-\frac{3}{2}u^{\prime}\right)\gamma^{4}+\frac{e^{-\frac{\phi}{4}}}{L}\cancel{D}_{S^{4}},\label{eq:3.29}
\end{equation}
where $\cancel{D}_{S^{4}}$ is the Dirac operator on a unit $S^{4}$.
Here the coordinates on the D7-brane are noted as $x^{a}=\left\{ t,x,y,x^{4}=\zeta,\Omega_{4}\right\} $
and $x^{\mu}=\left\{ t,x,y\right\} $. In addition, further calculations
lead to,

\begin{align}
\frac{1}{8}e^{\phi}F_{N}\Gamma^{\alpha}\Gamma^{N}\Gamma_{\alpha} & =-\frac{3}{4}\frac{u}{L}e^{\frac{3\phi}{2}}a\gamma^{2},\nonumber \\
\frac{e^{\phi}}{2\cdot8\cdot5!}\Gamma^{\alpha}F_{KLMNP}\Gamma^{KLMNP}\Gamma_{\alpha} & =\frac{1}{2}\frac{e^{-\frac{1}{4}\phi}}{L}\gamma^{56789},\nonumber \\
\frac{1}{2}\Gamma^{M}\partial_{M}\phi+\frac{1}{2}e^{\phi}F_{M}\Gamma^{M}\bar{\gamma} & =\frac{\zeta}{2L}e^{-\frac{\phi}{4}}\partial_{\zeta}\phi\gamma^{4}+e^{\frac{3\phi}{2}}\frac{u}{2L}a\gamma^{2}\bar{\gamma}.
\end{align}
Due to the kappa symmetry, the chirality of $\Psi$ can be fixed as
$\bar{\gamma}\Psi=\Psi$. And the ten-dimensional spinor $\Psi$ can
be further decomposed into a 3+1 dimensional part $\psi\left(x,\zeta\right)$
with holographic coordinate $\zeta$, an $S^{4}$ part $\varphi$
and a remaining two-dimensional part $\beta$ as,

\begin{equation}
\Psi=\frac{\left(2\pi\alpha^{\prime}\right)}{\sqrt{\mathcal{A}}}\psi\left(x,\zeta\right)\otimes\varphi\left(S^{4}\right)\otimes\beta,\label{eq:3.31}
\end{equation}
with $\beta^{\dagger}\beta=1$. The spinor $\varphi$ is the eigen
function on $S^{4}$ chosen by $\gamma^{56789}\Psi=\Psi$ satisfying
the Dirac equation on $S^{4}$ as \cite{Camporesi:1995fb},

\begin{equation}
\cancel{D}_{S^{4}}\varphi=\Lambda_{l}^{\pm}\varphi^{\pm l,s},\ \Lambda_{l}^{\pm}=\pm\left(2+l\right),l=0,1...,\label{eq:3.32}
\end{equation}
and normalization condition

\begin{equation}
\int\sqrt{g_{S^{4}}}d\Omega_{4}\varphi^{\pm l^{\prime},s^{\prime}\dagger}\varphi^{\pm l,s}=\delta^{l,l^{\prime}}\delta^{s,s^{\prime}},\label{eq:3.33}
\end{equation}
where $s,l$ refer to the quantum numbers of spherical harmonic function
on $S^{4}$ and $g_{S^{4}}$ denotes to the determinant of the metric
on a unit $S^{4}$. Inserting (\ref{eq:3.29}) - (\ref{eq:3.33})
into the fermionic action (\ref{eq:3.28}), it becomes the form as
$S_{\mathrm{D7}}^{f}=S_{f}^{\mathrm{k}}+S_{f}^{\mathrm{c}}$ given
by

\begin{align}
S_{f}^{\mathrm{k}}= & N_{c}\int d^{3}xd\zeta\bar{\psi}\left[\gamma^{\mu}\partial_{\mu}+\frac{\mathcal{B}}{\mathcal{A}}\gamma^{4}\partial_{4}+\left(\frac{\mathcal{C}}{\mathcal{A}}-\frac{\mathcal{A}^{\prime}\mathcal{B}}{2\mathcal{A}^{2}}\right)\gamma^{4}+\frac{\mathcal{D}}{\mathcal{A}}\right]\psi,\nonumber \\
S_{f}^{\mathrm{c}}= & N_{c}\int d^{3}xd\zeta\bar{\psi}\left(\frac{\mathcal{E}}{\mathcal{A}}\gamma^{2}\partial_{2}+\frac{\mathcal{F}}{\mathcal{A}}\gamma^{2}\right)\psi,\label{eq:3.34}
\end{align}
where the presented functions $\mathcal{A},\mathcal{B},\mathcal{C},\mathcal{D},\mathcal{E},\mathcal{F}$
are

\begin{align}
\mathcal{A} & =\frac{N_{c}L^{3}}{16\pi^{4}}\frac{e^{-\frac{\phi}{4}}}{u^{2}\zeta},\ \mathcal{B}=\frac{N_{c}L^{3}}{16\pi^{4}}\frac{e^{-\frac{\phi}{2}}}{u^{3}},\ \mathcal{C}=-\frac{N_{c}L^{3}}{16\pi^{4}}\frac{e^{-\frac{\phi}{2}}}{u^{4}}\left(\frac{1}{4}u\phi^{\prime}+\frac{3}{2}u^{\prime}\right),\nonumber \\
\mathcal{D} & =\frac{N_{c}L^{3}}{16\pi^{4}}\frac{e^{-\frac{\phi}{2}}}{u^{3}\zeta}\left(\Lambda_{l}^{\pm}-\frac{1}{2}\right),\ \mathcal{E}=\frac{N_{c}L^{3}}{16\pi^{4}}\frac{e^{-\frac{\phi}{4}}}{u^{2}\zeta}\frac{\hat{\phi}_{2}}{2}a^{2},\ \mathcal{F}=\frac{N_{c}L^{3}}{64\pi^{4}}\frac{e^{\frac{5\phi}{4}}}{u^{2}\zeta}a.
\end{align}
Choose the gamma matrices as 
\begin{equation}
\gamma^{\mu}=i\left(\begin{array}{cc}
0 & \sigma^{\mu}\\
\bar{\sigma}^{\mu} & 0
\end{array}\right),\gamma^{4}=\left(\begin{array}{cc}
1 & 0\\
0 & -1
\end{array}\right),
\end{equation}
where $\sigma^{\mu}=\left(1,-\tau^{i}\right),\bar{\sigma}^{\mu}=\left(1,\tau^{i}\right),i=1,2,3$
and $\tau^{i}$ is the Pauli matrix. Decomposing the spinor $\psi\left(x,\zeta\right)$
by using a series of completed functions $f_{\pm}^{\left(n\right)}$
as,

\begin{equation}
\psi=\sum_{n,m}\left(\begin{array}{c}
\psi_{+}^{\left(n\right)}f_{+}^{\left(n\right)}\\
\psi_{-}^{\left(m\right)}f_{-}^{\left(m\right)}
\end{array}\right),\label{eq:3.37}
\end{equation}
the actions presented in (\ref{eq:3.34}) becomes

\begin{align}
S_{f}^{\mathrm{k}}= & N_{c}\int d^{3}xd\zeta\sum_{n,m}\bigg\{-i\psi_{+}^{\left(n\right)\dagger}f_{+}^{\left(n\right)}\bar{\sigma}^{\mu}\partial_{\mu}\psi_{+}^{\left(m\right)}f_{+}^{\left(m\right)}-i\psi_{-}^{\left(m\right)\dagger}f_{-}^{\left(m\right)}\sigma^{\mu}\partial_{\mu}\psi_{-}^{\left(m\right)}f_{-}^{\left(m\right)}\nonumber \\
 & -\psi_{-}^{\left(m\right)\dagger}\psi_{+}^{\left(n\right)}f_{-}^{\left(m\right)}\left[\frac{\mathcal{B}}{\mathcal{A}}\partial_{4}f_{+}^{\left(n\right)}+\left(\frac{\mathcal{C}}{\mathcal{A}}-\frac{\mathcal{B}}{2\mathcal{A}^{2}}\frac{d\mathcal{A}}{d\zeta}\right)f_{+}^{\left(n\right)}+\frac{\mathcal{D}}{\mathcal{A}}f_{+}^{\left(n\right)}\right]\nonumber \\
 & -\psi_{+}^{\left(n\right)\dagger}\psi_{-}^{\left(m\right)}f_{+}^{\left(n\right)}\left[-\frac{\mathcal{B}}{\mathcal{A}}\partial_{4}f_{-}^{\left(m\right)}-\left(\frac{\mathcal{C}}{\mathcal{A}}-\frac{\mathcal{B}}{2\mathcal{A}^{2}}\frac{d\mathcal{A}}{d\zeta}\right)f_{-}^{\left(m\right)}+\frac{\mathcal{D}}{\mathcal{A}}f_{-}^{\left(m\right)}\right]\bigg\},\label{eq:3.38}
\end{align}
and

\begin{align}
S_{f}^{\mathrm{c}}= & -iN_{c}\int d^{3}xd\zeta\sum_{n,m}\left[\psi_{-}^{\left(m\right)\dagger}\sigma^{2}\left(\frac{\mathcal{E}}{\mathcal{A}}\partial_{2}+\frac{\mathcal{F}}{\mathcal{A}}\right)\psi_{-}^{\left(n\right)}+\psi_{+}^{\left(m\right)\dagger}\bar{\sigma}^{2}\left(\frac{\mathcal{E}}{\mathcal{A}}\partial_{2}+\frac{\mathcal{F}}{\mathcal{A}}\right)\psi_{+}^{\left(n\right)}\right]f_{+}^{\left(m\right)}f_{+}^{\left(n\right)}.\label{eq:3.39}
\end{align}
Imposing the normalization condition for the basis functions, 
\begin{figure}
\begin{centering}
\includegraphics[scale=0.38]{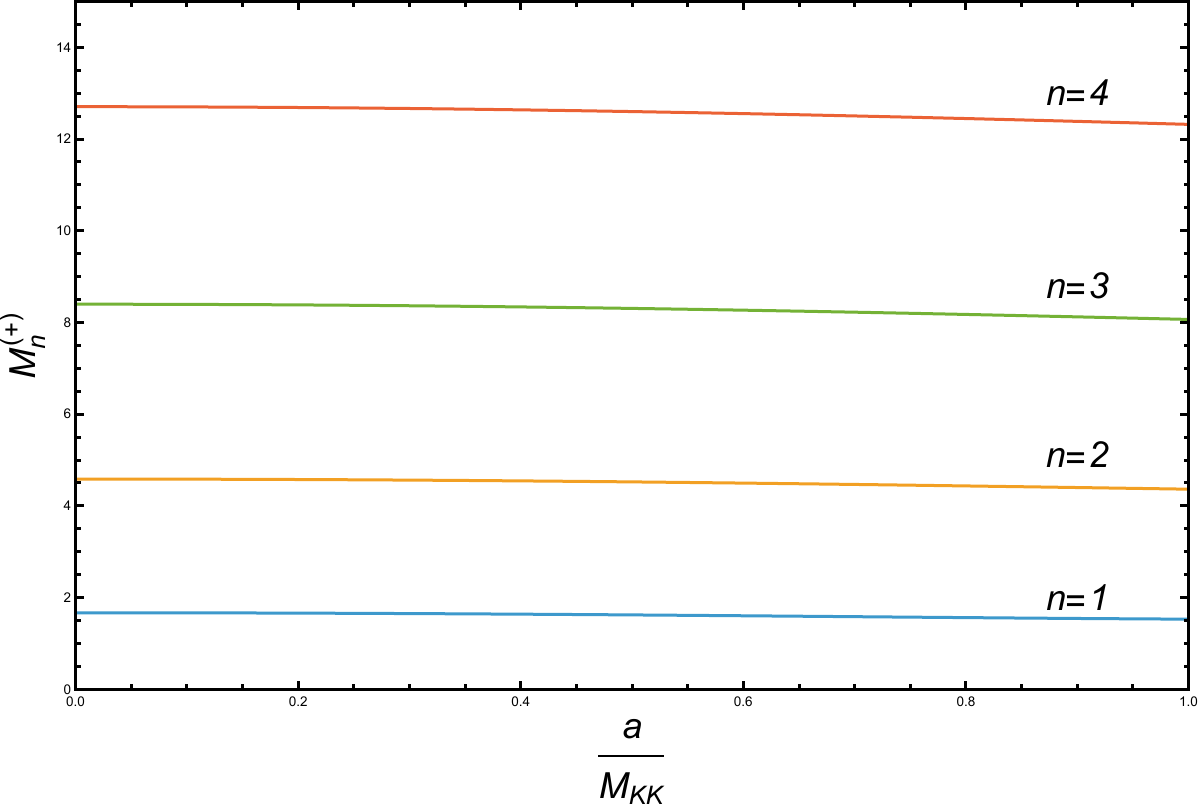}\includegraphics[scale=0.38]{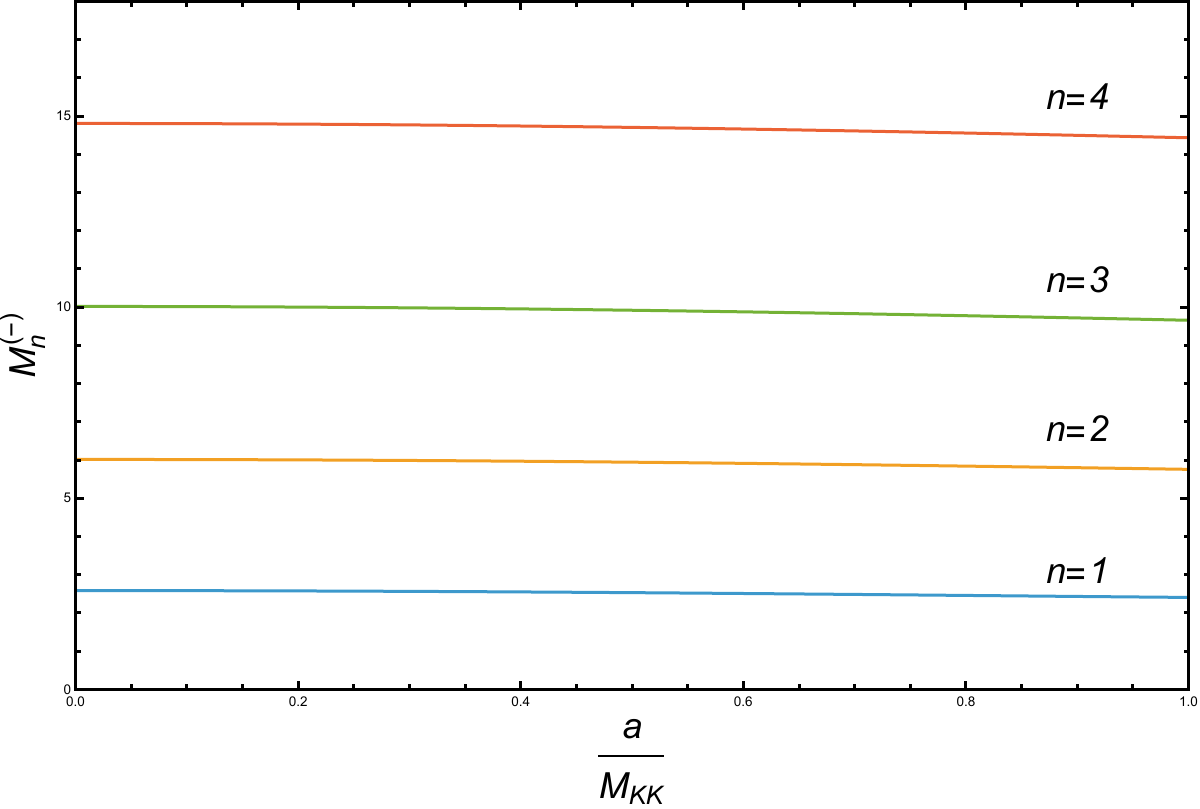}
\par\end{centering}
\caption{\label{fig:4}The mass spectrum $M_{n}^{\left(\pm\right)}$ of the
fermionic baryon as a function of $\frac{a}{M_{KK}}$.}
\end{figure}
 
\begin{table}
\begin{centering}
\begin{tabular}{|>{\centering}p{2.5cm}|>{\centering}p{1.5cm}|>{\centering}p{1.5cm}|>{\centering}p{1.5cm}|>{\centering}p{1.5cm}|>{\centering}p{1.5cm}|>{\centering}p{1.5cm}|}
\hline 
$n$ & $1$ & $2$ & $3$ & $4$ & $5$ & $6$\tabularnewline
\hline 
\hline 
$R_{T,+}^{\left(n\right)}\left(\times10^{-3}\right)$ & $-2.8$ & $-1.82$ & $-1.33$ & $-1.05$ & $-0.83$ & $-0.69$\tabularnewline
\hline 
$R_{T,-}^{\left(n\right)}\left(\times10^{-3}\right)$ & $-18.9$ & $-5.42$ & $-2.7$ & $-1.66$ & $-1.09$ & $-0.82$\tabularnewline
\hline 
\end{tabular}
\par\end{centering}
\begin{centering}
\begin{tabular}{|>{\centering}p{2.5cm}|>{\centering}p{1.5cm}|>{\centering}p{1.5cm}|>{\centering}p{1.5cm}|>{\centering}p{1.5cm}|>{\centering}p{1.5cm}|>{\centering}p{1.5cm}|}
\hline 
$\left(n,m\right)$ & $\left(1,2\right)$ & $\left(1,3\right)$ & $\left(1,4\right)$ & $\left(2,3\right)$ & $\left(2,4\right)$ & $\left(3,4\right)$\tabularnewline
\hline 
\hline 
$R_{S,+}^{\left(n,m\right)}\left(\times10^{-3}\right)$ & $-1.98$ & $1.46$ & $1.1$ & $1.43$ & $1.12$ & $-1.11$\tabularnewline
\hline 
$R_{S,-}^{\left(n,m\right)}\left(\times10^{-3}\right)$ & $-5.99$ & $-2.86$ & $1.4$ & $-2.98$ & $1.71$ & $1.79$\tabularnewline
\hline 
\end{tabular}
\par\end{centering}
\caption{\label{tab:6}The lowest dragging coefficients $R_{T,\pm}^{\left(n\right)},R_{S,\pm}^{\left(m,n\right)}$
presented in the baryonic action for $1\protect\leq m,n\protect\leq6$.}
\end{table}

\begin{equation}
M_{KK}\int d\zeta f_{\pm}^{\left(n\right)}f_{\pm}^{\left(m\right)}=\delta^{mn},\label{eq:3.40}
\end{equation}
with eigen equations ($M_{\left(n\right)}=\Lambda_{n}M_{KK}$)

\begin{align}
\frac{\mathcal{B}}{\mathcal{A}}\partial_{4}f_{+}^{\left(n\right)}+\left(\frac{\mathcal{C}}{\mathcal{A}}-\frac{\mathcal{B}}{2\mathcal{A}^{2}}\frac{d\mathcal{A}}{d\zeta}\right)f_{+}^{\left(n\right)}+\frac{\mathcal{D}}{\mathcal{A}}f_{+}^{\left(n\right)} & =M_{\left(n\right)}f_{-}^{\left(n\right)},\nonumber \\
-\frac{\mathcal{B}}{\mathcal{A}}\partial_{4}f_{-}^{\left(n\right)}-\left(\frac{\mathcal{C}}{\mathcal{A}}-\frac{\mathcal{B}}{2\mathcal{A}^{2}}\frac{d\mathcal{A}}{d\zeta}\right)f_{-}^{\left(n\right)}+\frac{\mathcal{D}}{\mathcal{A}}f_{-}^{\left(n\right)} & =M_{\left(n\right)}f_{+}^{\left(n\right)},\label{eq:3.41}
\end{align}
into action in (\ref{eq:3.38}), it becomes the standard kinetic terms
as,
\begin{equation}
S_{f}^{\mathrm{k}}=-\frac{N_{c}}{M_{KK}}\int d^{3}x\sum_{n}\left\{ i\psi_{+}^{\left(n\right)\dagger}\bar{\sigma}^{\mu}\partial_{\mu}\psi_{+}^{\left(n\right)}+i\psi_{-}^{\left(n\right)\dagger}\sigma^{\mu}\partial_{\mu}\psi_{-}^{\left(n\right)}+M_{\left(n\right)}\left[\psi_{-}^{\left(n\right)\dagger}\psi_{+}^{\left(n\right)}+\psi_{+}^{\left(n\right)\dagger}\psi_{-}^{\left(n\right)}\right]\right\} .\label{eq:3.42}
\end{equation}
And the action (\ref{eq:3.39}) becomes,

\begin{align}
S_{f}^{\mathrm{c}}= & -i\frac{N_{c}}{M_{KK}}\int d^{3}x\sum_{n}\bigg\{\psi_{-}^{\left(n\right)\dagger}\sigma^{2}\left[\frac{a^{2}}{M_{KK}^{2}}R_{T,-}^{\left(n\right)}\partial_{2}+\frac{a}{M_{KK}}\right]\psi_{-}^{\left(n\right)}\nonumber \\
 & +\psi_{+}^{\left(n\right)\dagger}\bar{\sigma}^{2}\left[\frac{a^{2}}{M_{KK}^{2}}R_{T,+}^{\left(n\right)}\partial_{2}+\frac{a}{M_{KK}}\right]\psi_{+}^{\left(n\right)}\bigg\}\nonumber \\
 & -i\frac{a^{2}N_{c}}{M_{KK}^{3}}\int d^{3}x\sum_{n\neq m}\left[R_{S,-}^{\left(nm\right)}\psi_{-}^{\left(n\right)\dagger}\sigma^{2}\partial_{2}\psi_{-}^{\left(m\right)}+R_{S,+}^{\left(nm\right)}\psi_{+}^{\left(n\right)\dagger}\bar{\sigma}^{2}\partial_{2}\psi_{+}^{\left(m\right)}\right],\label{eq:3.43}
\end{align}
with the associated coefficients

\begin{align}
\frac{M_{KK}^{3}}{a^{2}}\int d\zeta\frac{\mathcal{E}}{\mathcal{A}}f_{\pm}^{\left(m\right)}f_{\pm}^{\left(n\right)}\big|_{a=0} & =\begin{cases}
R_{T,\pm}^{\left(n\right)}, & m=n,\\
R_{S,\pm}^{\left(mn\right)}, & m\neq n.
\end{cases}\label{eq:3.44}
\end{align}
The mass spectrum $M_{\left(n\right)}$ of the fermionic baryon is
illustrated in Figure \ref{fig:4} and the coefficients given in (\ref{eq:3.44})
are numerically evaluated in Table \ref{tab:6}. According to the
numerical results, the spectrum of fermionic baryon is also suppressed
slightly when $a/M_{KK}$ increases. However, this fermionic system
does not include the imaginary frequency since all the coefficients
presented in (\ref{eq:3.42}) and (\ref{eq:3.43}) are real numbers.
This is distinct to the analysis of the bosonic mesons.

\section{The holographic hadronic interaction }

In this section, let us derive the lowest interaction terms involving
the holographic mesons and baryons. We will use the gauge of $A_{4}=0$
given in Section 3.1 for the vector boson.

\subsection{The interactions involving mesons}

\subsubsection*{Scalar meson}

\begin{figure}
\begin{centering}
\includegraphics[scale=0.5]{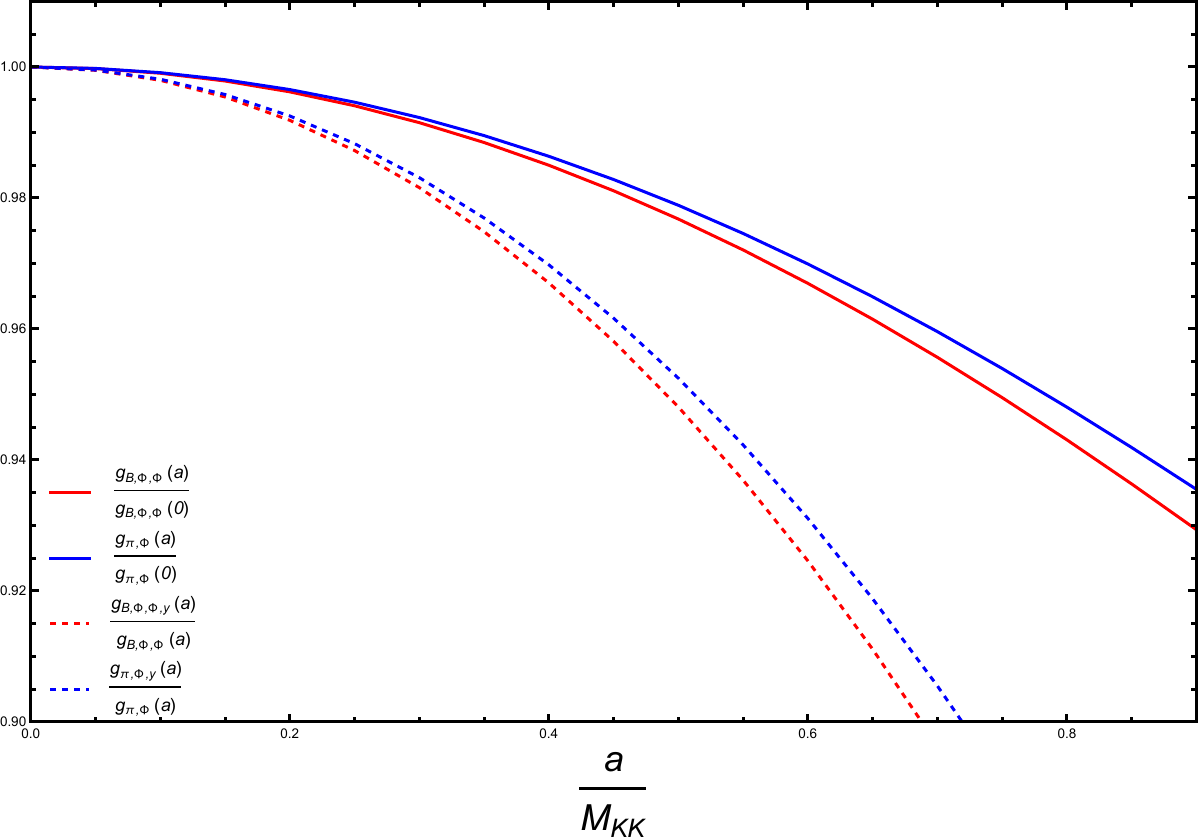}
\par\end{centering}
\caption{\label{fig:5}The ratios of the lowest coupling constants as functions
of $a/M_{KK}$ involving the scalar mesons $\Phi$. The lowest coupling
constants take the quantum numbers as $l,m,n=1$ for $g_{B,\Phi,\Phi}^{\left(l,n,m\right)},g_{\pi,\Phi}^{\left(n,m\right)}$
and $g_{B,\Phi,\Phi,y}^{\left(l,n,m\right)},g_{\pi,\Phi,y}^{\left(n,m\right)}$.}

\end{figure}

We start from the lowest action for the scalar meson given by

\begin{align}
S_{W} & =-\frac{1}{2}T_{7}\left(2\pi\alpha^{\prime}\right)^{2}\int_{\mathrm{D7}}d^{3}xd\zeta d\Omega_{4}e^{-\phi}\sqrt{-g_{\mathrm{D7}}}g^{ab}g_{ww}D_{a}WD_{b}W\nonumber \\
 & \equiv S_{\Phi}^{\mathrm{k}}+S_{\Phi}^{\mathrm{c}}+S_{B,\Phi,\pi}+S_{B,\Phi,\pi,y},\label{eq:4.1}
\end{align}
where $D_{a}=\partial_{a}W-i\left[A_{a},W\right]$ is the covariant
derivative operator for the adjoint scalar $W$. And $S_{\Phi}^{\mathrm{k}},S_{\Phi}^{\mathrm{c}}$
are the canonical kinetic plus dragging terms given in (\ref{eq:3.6})
(\ref{eq:3.7}). $S_{B,\Phi},S_{B,\Phi,y}$ are the lowest interaction
terms obtained from (\ref{eq:4.1}). By imposing the background metric
given in (\ref{eq:2.17}) with (\ref{eq:3.3}) (\ref{eq:3.27}), we
can derive the lowest interactions as

\begin{align}
S_{B,\Phi,\pi}= & \frac{i}{M_{KK}}\int d^{3}x\sum_{l,n,m}g_{B,\Phi,\Phi}^{\left(l,n,m\right)}\left[B_{\alpha}^{\left(l\right)},\Phi^{\left(n\right)}\right]\partial^{\alpha}\Phi^{\left(m\right)}\nonumber \\
 & -\frac{i}{M_{KK}}\int d^{3}x\sum_{n,m}g_{\pi,\Phi}^{\left(n,m\right)}\left[\partial_{\alpha}\pi,\Phi^{\left(n\right)}\right]\partial^{\alpha}\Phi^{\left(m\right)}+h.c.,\nonumber \\
S_{B,\Phi,\pi,y}= & \frac{i}{M_{KK}}\int d^{3}x\sum_{l,n,m}g_{B,\Phi,\Phi,y}^{\left(l,n,m\right)}\left[B_{y}^{\left(l\right)},\Phi^{\left(n\right)}\right]\partial_{y}\Phi^{\left(m\right)}\nonumber \\
 & -\frac{i}{M_{KK}}\int d^{3}x\sum_{n,m}g_{\pi,\Phi,y}^{\left(n,m\right)}\left[\partial_{y}\pi,\Phi^{\left(n\right)}\right]\partial_{y}\Phi^{\left(m\right)}+h.c.
\end{align}
And the presented coupling constants are obtained by

\begin{align}
g_{B,\Phi,\Phi}^{\left(l,n,m\right)} & =\frac{1}{2}\int d\zeta U_{s}b^{\left(l\right)}h^{\left(n\right)}h^{\left(m\right)},\ \ \ g_{\pi,\Phi}^{\left(n,m\right)}=g_{B,\Phi,\Phi}^{\left(0,n,m\right)},\nonumber \\
g_{B,\Phi,\Phi,y}^{\left(l,n,m\right)} & =\frac{1}{2}\int d\zeta e^{\phi}U_{s}b^{\left(l\right)}h^{\left(n\right)}h^{\left(m\right)},\ g_{\pi,\Phi,y}^{\left(n,m\right)}=g_{B,\Phi,\Phi,y}^{\left(0,n,m\right)}.\label{eq:4.3}
\end{align}
So the action (\ref{eq:4.1}) describes the interactions of the three-dimensional
scalar mesonic fields $\Phi^{\left(m\right)},\pi$ involving the vectors
$B_{\mu}^{\left(l\right)}$ in holography. And the interaction is
corrected by the anisotropy in the $y$ direction. Since our concern
is the dependence of the anisotropy and our holographic theory is
valid strictly in the large $N_{c}$ limit, we do not attempt to fit
the coupling constants presented in (\ref{eq:4.3}) to the experimental
data. Instead, we evaluate numerically the ratios $g\left(a\right)/g\left(0\right),g_{y}\left(a\right)/g\left(a\right)$
of the lowest coupling constants as functions of $a/M_{KK}$ to see
their trends with respect to the anisotropy, see them in Figure \ref{fig:5}.
Our numerical calculation displays that the lowest $g_{B,\Phi,\Phi}^{\left(1,1,1\right)},g_{\pi,\Phi}^{\left(1,1\right)}$
decrease as the anisotropic parameter $a/M_{KK}$ increases. Moreover,
$g_{B,\Phi,\Phi,y}^{\left(1,1,1\right)},g_{\pi,\Phi,y}^{\left(1,1\right)}$
decay rapidly than $g_{B,\Phi,\Phi}^{\left(1,1,1\right)},g_{\pi,\Phi}^{\left(1,1\right)}$
as $a/M_{KK}$ increases which implies the interactions involving
$B_{y}$ or mixed with $\partial_{y}$ becomes less important for
large $a/M_{KK}$ in this holographic setup. 

\subsubsection*{Vector meson}

\begin{figure}
\begin{centering}
\includegraphics[scale=0.38]{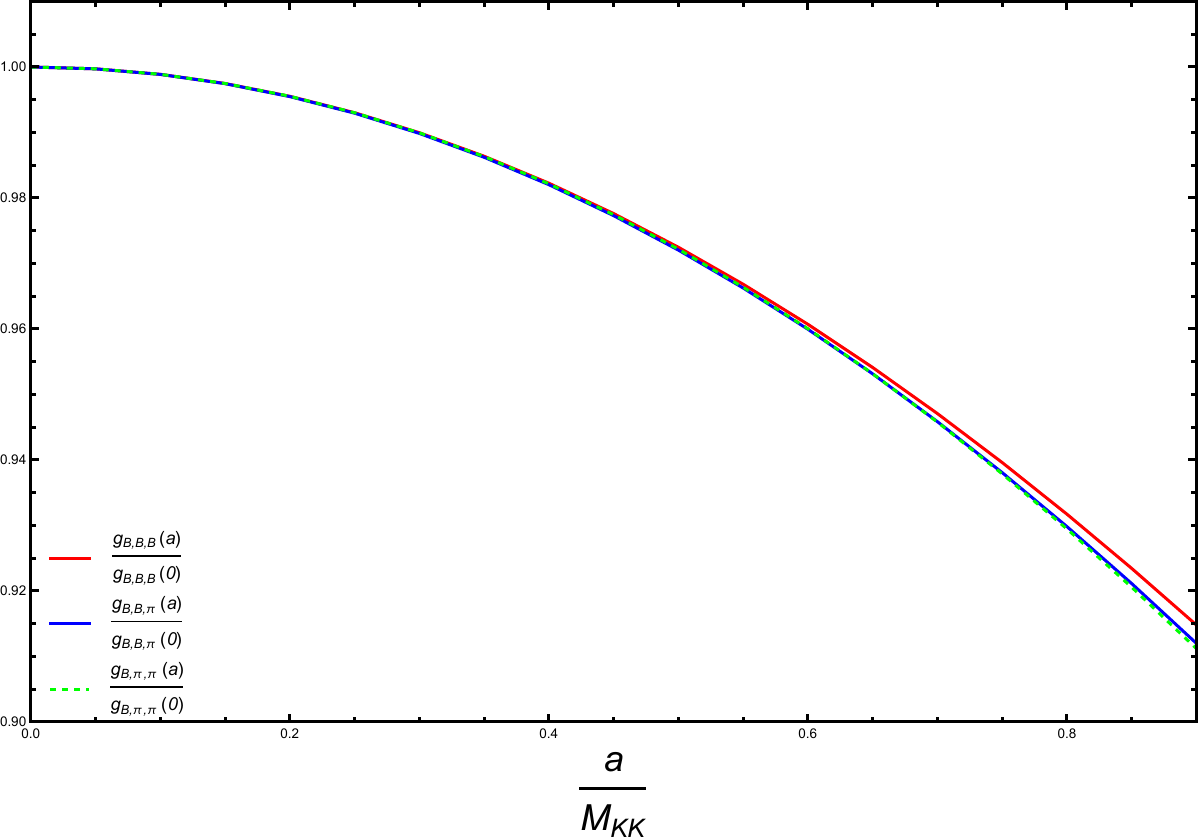}\includegraphics[scale=0.38]{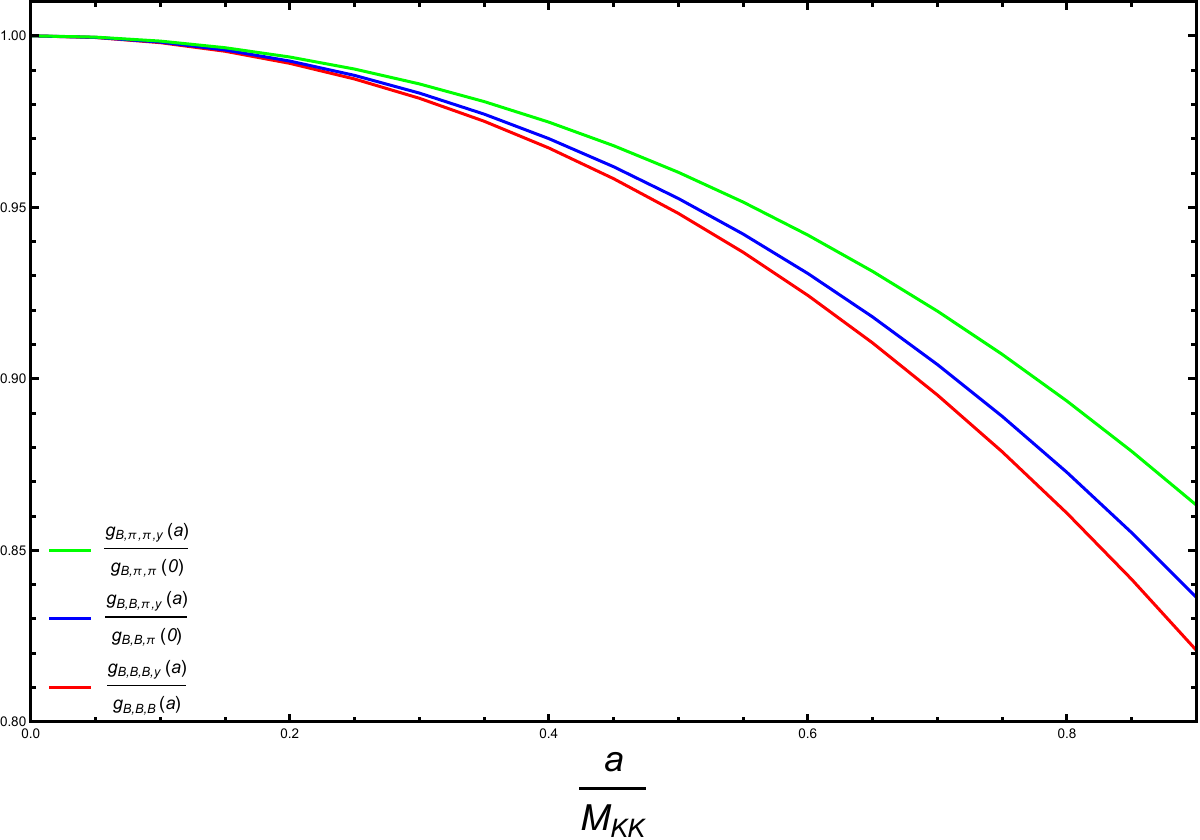}
\par\end{centering}
\caption{\label{fig:6}The ratios of the lowest coupling constants as functions
of $a/M_{KK}$ involving the vector mesons $B_{\mu}$. The lowest
coupling constants take the quantum numbers as $l,m,n=1$ for $g_{B,B,B}^{\left(l,m,n\right)},g_{B,B,\pi}^{\left(l,n\right)},g_{B,\pi,\pi}^{\left(l\right)}$
and $g_{B,B,B,y}^{\left(l,m,n\right)},g_{B,B,\pi,y}^{\left(l,n\right)},g_{B,\pi,\pi,y}^{\left(l\right)}$.}

\end{figure}

For the vector meson, while the derivation is a little lengthy, all
the interaction terms come from the D7-brane action involving the
gauge field as,

\begin{align}
S_{V}= & -\frac{1}{4}\left(2\pi\alpha^{\prime}\right)^{2}T_{\mathrm{D7}}\int_{\mathrm{D7}}d^{3}xd\zeta d\Omega_{4}e^{-\phi}\sqrt{-g_{\mathrm{D7}}}g^{ac}g^{bd}f_{ab}f_{cd}\nonumber \\
\equiv & S_{V}^{\mathrm{k}}+S_{V}^{\mathrm{c}}+S_{B,\pi}+S_{B,\pi,y},\label{eq:4.4}
\end{align}
where $S_{V}^{\mathrm{k}},S_{V}^{\mathrm{c}}$ are the canonical kinetic
action plus the dragging terms given in (\ref{eq:3.20}). The other
terms presented in action (\ref{eq:4.4}) can be derived as follows
by taking into account the background metric (\ref{eq:2.17}) and
(\ref{eq:3.27}),

\begin{align}
S_{B,\pi}= & -\frac{i}{M_{KK}}\int d^{3}x\bigg\{\sum_{l,m,n}g_{B,B,B}^{\left(l,m,n\right)}F_{\alpha\beta}^{\left(l\right)}\left[B^{\alpha\left(m\right)},B^{\beta\left(n\right)}\right]-\frac{1}{M_{KK}}\sum_{l,n}g_{B,B,\pi}^{\left(l,n\right)}F_{\alpha\beta}^{\left(l\right)}\left[\partial^{\alpha}\pi,B^{\beta\left(n\right)}\right]\nonumber \\
 & +\frac{1}{M_{KK}^{2}}\sum_{l}g_{B,\pi,\pi}^{\left(l\right)}F_{\alpha\beta}^{\left(l\right)}\left[\partial^{\alpha}\pi,\partial^{\beta}\pi\right]\bigg\}+h.c.,\nonumber \\
S_{B,\pi,y}= & -\frac{i}{M_{KK}}\int d^{3}x\bigg\{\sum_{l,m,n}F_{\alpha y}^{\left(l\right)}\left[B^{\alpha\left(m\right)},B^{y\left(n\right)}\right]g_{B,B,B,y}^{\left(l,m,n\right)}-\frac{1}{M_{KK}}\sum_{l,n}g_{B,B,\pi,y}^{\left(l,n\right)}F_{\alpha y}\left[\partial^{\alpha}\pi,B^{y\left(n\right)}\right]\nonumber \\
 & +\frac{1}{M_{KK}^{2}}\sum_{l}g_{B,\pi,\pi,y}^{\left(l\right)}F_{\alpha y}^{\left(l\right)}\left[\partial^{\alpha}\pi,\partial^{y}\pi\right]\bigg\}+h.c.
\end{align}
and the all presented coupling constants are given by

\begin{align}
g_{B,B,B}^{\left(l,m,n\right)} & =\frac{1}{2}\int d\zeta U_{v}b^{\left(l\right)}b^{\left(m\right)}b^{\left(n\right)},\ g_{B,B,\pi}^{\left(l,n\right)}=2g_{B,B,B}^{\left(0,l,,n\right)},\ g_{B,\pi,\pi}^{\left(l\right)}=g_{B,B,B}^{\left(0,0,l\right)},\nonumber \\
g_{B,B,B,y}^{\left(l,m,n\right)} & =\frac{1}{2}\int d\zeta U_{v}e^{\phi}b^{\left(l\right)}b^{\left(m\right)}b^{\left(n\right)},\ g_{B,B,\pi,y}^{\left(l,n\right)}=2g_{B,B,B,y}^{\left(0,l,,n\right)},\ g_{B,\pi,\pi}^{\left(l\right)}=g_{B,B,B,y}^{\left(0,0,l\right)},\label{eq:4.6}
\end{align}
Thus, the action (\ref{eq:4.4}) describes the interaction of the
vectors $B_{\mu}^{\left(l\right)}$ involving the scalars $\pi$.
We also evaluate numerically the ratios of the lowest coupling constants
given in (\ref{eq:4.6}) as functions of $a/M_{KK}$ which are illustrated
in Figure \ref{fig:6}. 
The numerical results show that the behavior of the lowest coupling
constants in (\ref{eq:4.6}) is basically similar as the case of the
scalar meson. Hence it is a parallel verification of that the bosonic
system may include unstable modes.

\subsection{The interactions involving baryons}

In order to take into account the lowest interactions of fermionic
baryons and bosonic mesons, we need to expand the fermionic action
for the D7-branes up to the leading order of the gauge field strength
$f_{ab}$. Since the full action involving the worldvolume fermions
and bosons is given in Appendix B, its leading order can be obtained
as,

\begin{align}
S_{f} & =S_{f}^{\mathrm{k}}+S_{f}^{\mathrm{c}}+S_{f}^{\left(1\right)}+S_{f}^{\left(2\right)},\nonumber \\
S_{f}^{\left(1\right)} & =-i\frac{T_{7}}{2}\left(2\pi\alpha^{\prime}\right)\int_{\mathrm{D7}}d^{8}x\sqrt{-g}e^{-\phi}\bar{\Psi}\frac{1}{2}\gamma^{3}\Gamma^{ab}f_{ab}\left(\Gamma^{c}\hat{D}_{c}-\varDelta\right)\Psi,\nonumber \\
S_{f}^{\left(2\right)} & =-i\frac{T_{7}}{2}\left(2\pi\alpha^{\prime}\right)\int_{\mathrm{D7}}d^{8}x\sqrt{-g}e^{-\phi}\bar{\Psi}\left(1-\gamma^{3}\right)\gamma^{3}\bar{\gamma}\Gamma^{a}f_{ab}g^{bc}\hat{D}_{c}\Psi.
\end{align}
By imposing the decomposition (\ref{eq:3.28}) - (\ref{eq:3.33}),
$S_{f}^{\left(1\right)}$ becomes vanished since it contains a term
of $\left(\Gamma^{c}\hat{D}_{c}-\varDelta\right)\Psi$ which is the
Dirac equation or the onshell condition for the functions $f_{\pm}^{\left(n\right)}$
exactly. Nevertheless, $S_{f}^{\left(2\right)}$ can be written in
terms of a very lengthy sum as,

\begin{equation}
S_{f}^{\left(2\right)}=\sum_{t=1}^{3}S_{B,\psi}^{\left(t\right)}+\sum_{t=1}^{4}S_{B,\psi,\pi}^{\left(t\right)},\label{eq:4.8}
\end{equation}
which includes various interactions of vector mesons $B_{\mu}^{\left(n\right)}$,
scalar meson $\pi$ and spinor baryons $\psi_{\pm}^{\left(n\right)}$
and their corrections of derivatives with respect to the $y$ direction.
The full formulas of (\ref{eq:4.8}) can be found in Appendix C which
is too lengthy to be given here. To evaluate the dependence of the
anisotropy in the coupling constants, we pick up the characteristic
structures of the couplings as,

\begin{align}
g_{B,\psi,\left(r,s\right)}^{\left(n,l,m\right)} & =\int d\zeta Xb^{\left(n\right)}f_{s}^{\left(l\right)}f_{r}^{\left(m\right)},\ g_{B,\psi,y,\left(r,s\right)}^{\left(n,l,m\right)}=\int d\zeta Xe^{\phi}b^{\left(n\right)}f_{s}^{\left(l\right)}f_{r}^{\left(m\right)},\nonumber \\
g_{\psi,\pi,\left(r,s\right)}^{\left(m,n\right)} & =\int d\zeta X\frac{\zeta}{u}e^{-\frac{\phi}{4}}d^{\left(0\right)}f_{s}^{\left(m\right)}f_{r}^{\left(n\right)},\ g_{\psi,\pi,y,\left(r,s\right)}^{\left(m,n\right)}=\int d\zeta X\frac{\zeta}{u}e^{3\phi/4}d^{\left(0\right)}f_{s}^{\left(m\right)}f_{r}^{\left(n\right)},\label{eq:4.9}
\end{align}
where the indices ``$r,s$'' run over ``$+,-$'' of the basis
functions $f_{\pm}$ according to (\ref{eq:3.37}). 
\begin{figure}
\begin{centering}
\includegraphics[scale=0.38]{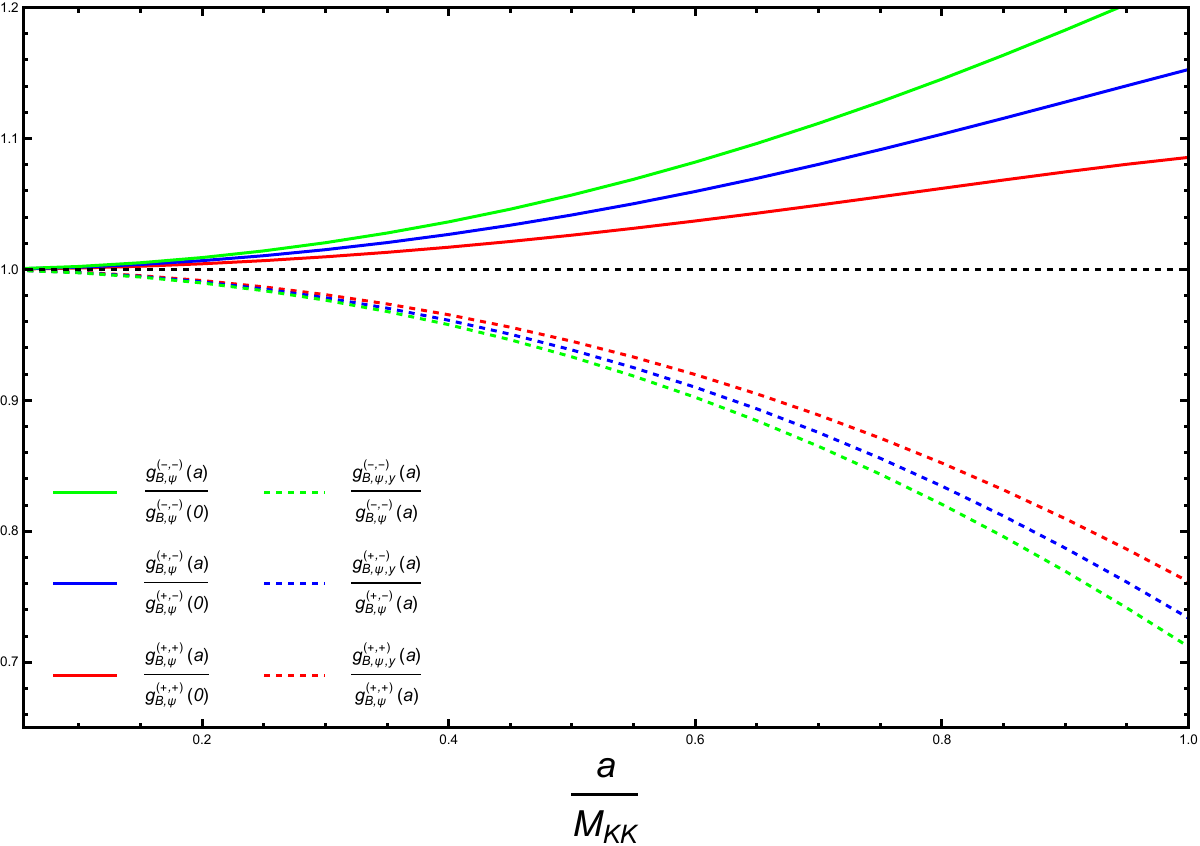}\includegraphics[scale=0.38]{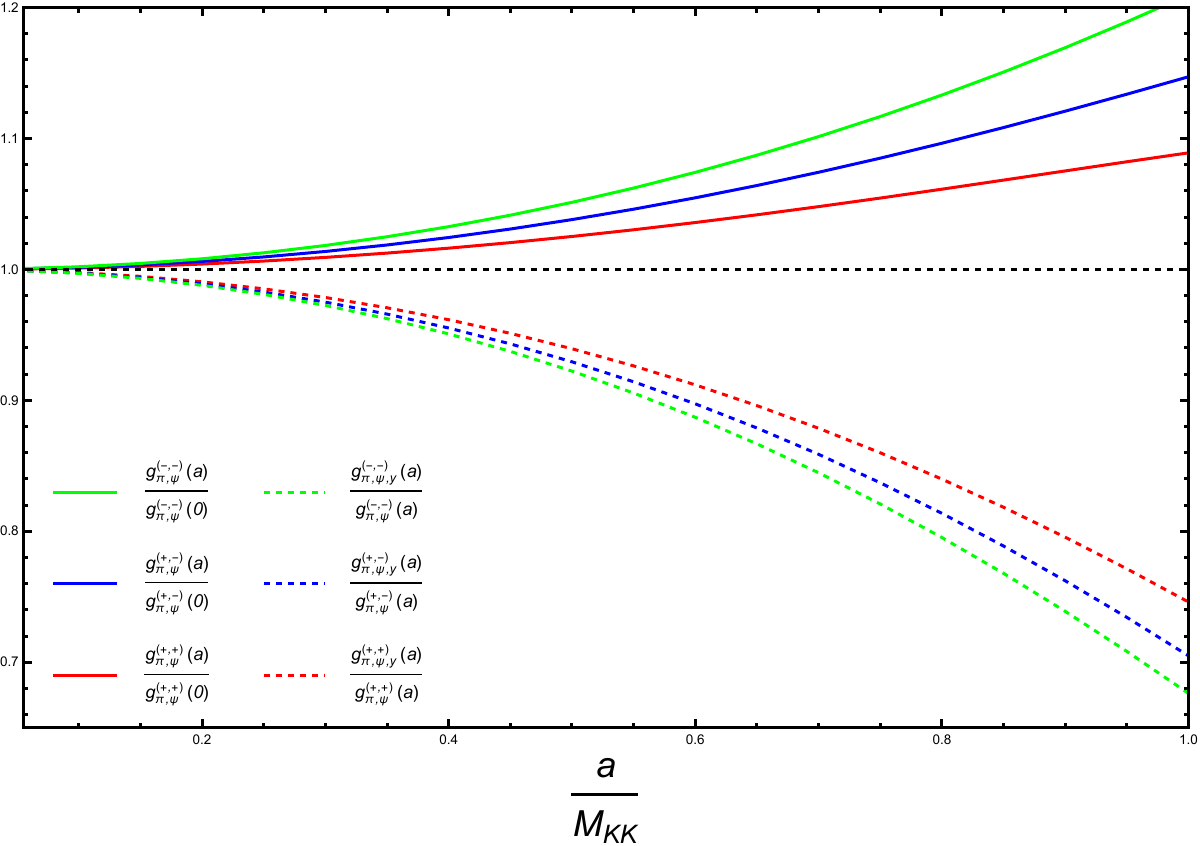}
\par\end{centering}
\caption{\label{fig:7}The ratios of the lowest coupling constants as functions
of $a/M_{KK}$ involving the spinor baryons $\psi$. The lowest coupling
constants take the quantum numbers as $l,m,n=1$ for $g_{B,\psi,\left(r,s\right)}^{\left(n,l,m\right)},g_{\psi,\pi,\left(r,s\right)}^{\left(m,n\right)}$
and $g_{B,\psi,y,\left(r,s\right)}^{\left(n,l,m\right)},g_{\psi,\pi,y,\left(r,s\right)}^{\left(m,n\right)}$.
The ``$+,-$'' refers to the value of the indices $r,s$ presented
in (\ref{eq:4.9}).}

\end{figure}
The coupling constants in (\ref{eq:4.9}) refers to the interaction
vertices as $\sim f_{\alpha\beta}^{\left(l\right)}\psi_{r}^{\left(m\right)\dagger}\sigma^{\beta}\partial^{\alpha}\psi_{s}^{\left(n\right)},\partial_{\alpha}\pi\eta^{\alpha\beta}\psi_{r}^{\left(m\right)\dagger}\partial_{\beta}\psi_{s}^{\left(n\right)}$
and their associated corrections with respect to $\sim f_{2\beta}^{\left(l\right)}\psi_{r}^{\left(m\right)\dagger}\sigma^{2}\partial^{\beta}\psi_{s}^{\left(n\right)},\partial_{y}\pi\psi_{r}^{\left(m\right)\dagger}\partial_{y}\psi_{s}^{\left(n\right)}$.
The numerical evaluation with respect to the lowest coupling constants
given in (\ref{eq:4.9}) is illustrated in Figure \ref{fig:7}. We
find while the interactions mixed with $f_{2\beta}^{\left(l\right)}$
or $\partial_{y}\pi$ become less important for large $a/M_{KK}$,
the coupling constant $g_{B,\psi,\left(r,s\right)}^{\left(1,1,1\right)},g_{\psi,\pi,\left(r,s\right)}^{\left(1,1\right)}$
increases for large $a/M_{KK}$. This behavior is opposite to the
analysis of the bosonic coupling constants due to the form of the
Dirac action. Altogether, as the anisotropic parameter $a/M_{KK}$
increases, all the fermionic modes are stable participating the hadronic
interactions while parts of bosonic modes becomes unstable decayed
off the hadronic interactions. Therefore, the hadronic interactions
involving fermion dominates in our holographic hadronic system for
large $a/M_{KK}$.

\section{Summary and discussion}

In this work, we investigate systematically the low-energy hadronic
physics by using the anisotropic D3-brane solution in holography.
We first construct the black D3-brane solution to be a confining geometry
then take into account the embedding of the flavor D7-branes and the
baryon vertex as a D5-brane. Afterwards the mass spectra and dragging
coefficients are numerically evaluated by deriving the canonical kinetic
actions for meson and baryons. These numerical results indicate that,
while the hadronic system is stable for $a/M_{KK}\ll1$, the bosonic
mesons may include unstable modes if the anisotropic parameter $a/M_{KK}$
becomes sufficiently large. Besides, we also evaluate the lowest coupling
constants numerically by analyzing the lowest interaction terms in
the concerned actions which implies that the fermionic baryon may
dominate the hadronic interactions if the anisotropic parameter $a/M_{KK}$
increases. 

With these in mind, we give our discussion of this work as follows.
First, this holographic approach illustrates that the dragging terms
presented in the effective actions are very necessary for an anisotropic
field theory since the dragging terms contribute additional dissipation
to affect the transport properties of the media. Qualitatively, according
to the AdS/CFT correspondence, the shear viscosity is $\eta\sim FT$
where $F$ is the drag force \cite{Chernicoff:2012iq,Son:2002sd},
and the bosonic dragging coefficients contribute to $F$ as $\mathcal{O}\left(a^{4}\right)$
in the expansion of $a\ll1$ according to this work. Thus, for $a\ll1$,
the shear viscosity can be decomposed as $\eta=\eta^{\left(iso\right)}+\delta\eta$
where $\eta^{\left(iso\right)}$ is the isometric part and $\delta\eta\sim a^{4}T$
is the contribution from the anisotropy. Picking up the entropy density
$s$ proportional to $\left(1+a^{2}\right)T$ \cite{Mateos:2011ix},
we find $\frac{\eta}{s}\sim\frac{1+a^{4}}{1+a^{2}}$ for $a\ll1$
based on this work, therefore the presented anisotropic dragging terms
clarify why the $\frac{\eta}{s}$ can break through the lower bound
$\frac{1}{4\pi}$ in this holographic approach\cite{Mateos:2011ix}.

Second, as the presented anisotropic parameter leads to an anisotropic
instanton configuration or a Chern-Simons term in the dual theory,
it would be interesting to compare this work with the the analysis
of the isotropic instantons or the isotropic Chern-Simons term in
the D3/D7 approach \cite{Li:2025ahp,Li:2024jkd,Li:2024apc}. As we
can see, the mass spectra of the various hadrons in this approach
trend to be reduced in the presence of the anisotropic instantons
and the interaction coupling constants for the bosonic mesons behave
similarly. Since these behaviors are different from the analysis of
the isotropic instantons studied in \cite{Li:2025ahp,Li:2024jkd},
it may imply the origin of the instability in the hadronic system
comes form the anisotropy itself, especially for that the anisotropy
$a$ is much larger than confinement energy scale $M_{KK}$ predicted
in \cite{Li:2022wwv}. And this conclusion for the low-energy hadronic
physics in this approach is additional to the framework \cite{Mateos:2011ix}.

Last but not least, we do not attempt to fit the mass spectra and
coupling constants in this work to the experimental data since our
theory is three-dimensional. Instead, we focus on the trends or the
dependence with respect to the anisotropy which should be less depended
on the dimension. In this sense, this work should also be valid to
four-dimensional QCD with respect to the dependence on the anisotropy.

\section*{Acknowledgements}

We would like to thank Xin-li Sheng and Prof. Yan-qing Zhao for helpful
discussion. This work is supported by the National Natural Science
Foundation of China (NSFC) under Grant No. 12005033 and the Fundamental
Research Funds for the Central Universities under Grant No. 3132026198.

\section{Appendixes}

\subsection*{Appendix A: The analytical functions for the anisotropic background}

The functions $\mathcal{F},\mathcal{B},\phi$ presented in the black
brane background (\ref{eq:2.2}) in the high temperature limit $T\rightarrow\infty$
(i.e. $\delta t_{E}\rightarrow0$) can be written in terms of a series
of $au_{T}$, up to $\mathcal{O}\left(a^{2}\right)$ they are \cite{Avila:2016mno,Li:2022wwv,Mateos:2011tv},

\begin{align}
\mathcal{F}\left(u\right) & =1-\frac{u^{4}}{u_{H}^{4}}+a^{2}\hat{\mathcal{F}}_{2}\left(u\right)+\mathcal{O}\left(a^{4}\right),\nonumber \\
\mathcal{B}\left(u\right) & =1+a^{2}\hat{\mathcal{B}}_{2}\left(u\right)+\mathcal{O}\left(a^{4}\right),\nonumber \\
\phi\left(u\right) & =a^{2}\hat{\phi}_{2}\left(u\right)+\mathcal{O}\left(a^{4}\right),\tag{A-1}\label{eq:A-1}
\end{align}
where

\begin{align}
\hat{\mathcal{F}}_{2}\left(u\right) & =\frac{1}{24u_{H}^{2}}\left[8u^{2}\left(u_{H}^{2}-u^{2}\right)-10u^{4}\log2+\left(3u_{H}^{4}+7u^{4}\right)\log\left(1+\frac{u^{2}}{u_{H}^{2}}\right)\right],\nonumber \\
\hat{\mathcal{B}}_{2}\left(u\right) & =-\frac{u_{H}^{2}}{24}\left[\frac{10u^{2}}{u_{H}^{2}+u^{2}}+\log\left(1+\frac{u^{2}}{u_{H}^{2}}\right)\right],\nonumber \\
\hat{\phi}_{2}\left(u\right) & =-\frac{u_{H}^{2}}{4}\log\left(1+\frac{u^{2}}{u_{H}^{2}}\right).\tag{A-2}\label{eq:A-2}
\end{align}
Note that (\ref{eq:A-1}) and (\ref{eq:A-2}) can be further equivalently
written as a series of $a/T$ once (\ref{eq:2.6}) is imposed. In
this sense, the high temperature limit refers to the case $T\gg a$
exactly in \cite{Mateos:2011tv} which corresponds to $u_{H}a\ll1$
or $a/T\ll1$. In the black brane background (\ref{eq:2.2}) , we
use $u_{H}$ to refer to the horizon position, and in the bubble background
(\ref{eq:2.7}), $u_{H}$ is replaced by $u_{KK}$representing the
bottom of the bulk spacetime since the double wick rotation induces
to the replacement $T\rightarrow M_{KK}/\left(2\pi\right),\delta t_{E}\rightarrow\delta z$
in the black brane solution. Note that the formulas of $\mathcal{F},\mathcal{B},\phi$
remain while we replace $u_{H}$ by $u_{KK}$. Accordingly, the high
temperature limit in the black brane solution corresponds to the limit
of dimension reduction in the bubble solution which means the compactified
direction $z$ shrinks to zero in (\ref{eq:2.7}) (i.e. $\delta z\rightarrow0$),
or equivalently $M_{KK}\rightarrow\infty$. Therefore the dual theory
in this limit trend to be purely three-dimensional.

\subsection*{Appendix B: The action for the worldvolume fields on the D-brane}

The dynamics of coincident $N$ D$p$-branes with non-Abelian excitations
can be described by the D-brane action. In principle, the D-brane
action could be obtained by using T-duality as the standard technique
in string theory. Let us consider a stack of $N$ D$p$-branes in
the $D$ dimensional spacetime parametrized by $\left\{ X^{M}\right\} ,M=0,1...D-1$
without loss of generality. Note the indices are defined as $\alpha,\beta=0,1...p$
and $i,j,k=p+1...D-1$ which denote respectively the directions parallel
and vertical to the $N$ D$p$-branes. The full action of such $N$
D$p$-branes is collected as,

\[
S_{\mathrm{D}_{p}}=S_{b}+S_{f},\tag{B-1}
\]
where the bosonic part $S_{b}$ is consisted of a DBI (Dirac-Born-Infeld)
plus a WZ (Wess-Zumino) term \cite{Johnson:2002} as

\begin{align}
S_{b}= & S_{\mathrm{DBI}}+S_{\mathrm{WZ}}\nonumber \\
S_{\mathrm{DBI}}= & -NT_{\mathrm{D}_{p}}\mathrm{STr}\int_{\mathrm{D}_{p}}d^{p+1}\xi e^{-\phi}\sqrt{\det\left[Q_{\ j}^{i}\right]\det\left\{ -\left[E_{\alpha\beta}+E_{\alpha i}\left(Q^{-1}-\delta\right)^{ij}E_{j\beta}+2\pi\alpha^{\prime}f_{\alpha\beta}\right]\right\} },\nonumber \\
S_{\mathrm{WZ}}= & N\mu_{p}\sum_{n=0,1...}\int_{\mathrm{D}_{p}}C_{p-2n+1}\wedge\frac{\left(B+2\pi\alpha^{\prime}f\right)^{n}}{n!},\nonumber \\
Q_{\ j}^{i}= & \delta^{i}\ j+2\pi\alpha^{\prime}\left[\varphi^{i},\varphi^{k}\right]E_{kj},\ E_{MN}=g_{MN}+B_{MN}.\tag{B-2}\label{eq:B-2}
\end{align}
The $T_{\mathrm{D}_{p}}=\frac{1}{\left(2\pi\right)^{p}l_{s}^{p+1}}$
is the tension of the D$p$-brane, $\phi$ is the bulk dilaton. Here
we use $g_{\alpha\beta},B_{\alpha\beta}$ to denote the induced metric
of the $D$ dimensional spacetime and the induced 2-form field respectively.
The ``STr'' represents the ``symmetric trace'' and $f$ is the
non-Abelian gauge field strength defined as $f_{\alpha\beta}=\partial_{\alpha}A_{\beta}-\partial_{\beta}A_{\alpha}-i\left[A_{\alpha},A_{\beta}\right]$.
$\varphi^{i}$ 's refer to the transverse adjoint modes of the D$p$-branes
which are the ``T-dualitized'' coordinates given by $2\pi\alpha^{\prime}\varphi^{i}=X^{i}$.
We have chosen the ``static gauge'' throughout the manuscript to
gauge away the 2-form field $B$ i.e. $B_{MN}=0$. With all of these
in hand, the DBI action in (\ref{eq:B-2}) could be expanded up to
quadratic order as,

\begin{align}
S_{\mathrm{DBI}}= & -T_{\mathrm{D}_{p}}\left(2\pi\alpha^{\prime}\right)^{2}\mathrm{Tr}\int d^{p+1}\xi e^{-\phi}\sqrt{-g}\left[1+\frac{1}{4}f_{\alpha\beta}f^{\alpha\beta}+\frac{1}{2}D_{\alpha}\varphi^{i}D^{\alpha}\varphi^{i}-\frac{1}{4}\left[\varphi^{i},\varphi^{j}\right]^{2}\right]\nonumber \\
 & +\mathrm{high\ orders},\tag{B-3}\label{eq:B-3}
\end{align}
where the covariant derivative is given as $D_{\alpha}\varphi^{i}=\partial_{\alpha}\varphi^{i}-i\left[A_{\alpha},\varphi^{i}\right]$. 

There is in addition a fermionic part $S_{f}$ of the D-brane action
which is collected as \cite{Li:2025ahp,Li:2023wyb,Nakas:2020hyo,Marolf:2003ye,Marolf:2003vf,Li:2024jkd} 

\begin{equation}
S_{f}=i\frac{NT_{\mathrm{D}_{p}}}{2}\int_{\mathrm{D}_{p}}d^{p+1}xe^{-\phi}\sqrt{-\left(g+f\right)}\bar{\Psi}\left(1-\Gamma_{\mathrm{D}_{p}}\right)\left(\Gamma^{\alpha}\hat{D}_{\alpha}-\Delta+\mathrm{L}_{\mathrm{D}_{p}}\right)\Psi,\tag{B-4}\label{eq:B-4}
\end{equation}
where $\Psi$ is the worldvolume fermion on the $N$ D$p$-branes.
For the type IIA string theory ($p$ is even number), the presented
operators are,

\begin{align}
\hat{D}_{\alpha} & =\nabla_{\alpha}+\frac{1}{4\cdot2!}H_{\alpha NK}\Gamma^{NK}\bar{\gamma}+\frac{1}{8}e^{\phi}\left(\frac{1}{2!}F_{NK}\Gamma^{NK}\Gamma_{\alpha}\bar{\gamma}+\frac{1}{4!}F_{KLNP}\Gamma^{KLNP}\Gamma_{\alpha}\right),\nonumber \\
\Delta & =\frac{1}{2}\left(\Gamma^{M}\partial_{M}\phi+\frac{1}{2\cdot3!}H_{MNK}\Gamma^{MNK}\bar{\gamma}\right)+\frac{1}{8}e^{\phi}\left(\frac{3}{2!}F_{MN}\Gamma^{MN}\bar{\gamma}+\frac{1}{4!}F_{KLNP}\Gamma^{KLNP}\right),\nonumber \\
\Gamma_{\mathrm{D}_{p}} & =\frac{1}{\sqrt{-\left(g+\mathcal{F}\right)}}\sum_{q}\frac{\epsilon^{\alpha_{1}...\alpha_{2q}\beta_{1}...\beta_{p-2q+1}}}{q!2^{q}\left(p-2q+1\right)!}\mathcal{F}_{\alpha_{1}\alpha_{2}}...\mathcal{F}_{\alpha_{2q-1}\alpha_{2q}}\Gamma_{\beta_{1}...\beta_{p-2q+1}}\bar{\gamma}^{\frac{p-2q+2}{2}},\nonumber \\
\mathrm{L}_{\mathrm{D}_{p}} & =\sum_{q}\frac{\epsilon^{\alpha_{1}...\alpha_{2q}\beta_{1}...\beta_{p-2q+1}}}{q!2^{q}\left(p-2q+1\right)!}\frac{\left(-\bar{\gamma}\right)^{\frac{p}{2}-q+1}}{\sqrt{-\left(g+\mathcal{F}\right)}}\mathcal{F}_{\alpha_{1}\alpha_{2}}...\mathcal{F}_{\alpha_{2q-1}\alpha_{2q}}\Gamma_{\beta_{1}...\beta_{p-2q+1}}^{\ \ \ \ \ \ \ \ \ \ \ \ \ \lambda}\hat{D}_{\lambda}.\tag{B-5}\label{eq:B-5}
\end{align}
For the type IIB string theory ($p$ is odd number), the presented
operators are,

\begin{align}
\hat{D}_{\alpha}= & \nabla_{\alpha}+\frac{1}{4\cdot2!}H_{\alpha NK}\Gamma^{NK}\bar{\gamma}-\frac{1}{8}e^{\phi}\big(F_{N}\Gamma^{N}\Gamma_{\alpha}\bar{\gamma}+\frac{1}{3!}F_{KLN}\Gamma^{KLN}\Gamma_{\alpha}\nonumber \\
 & +\frac{1}{2\cdot5!}F_{KLMNP}\Gamma^{KLMNP}\Gamma_{\alpha}\bar{\gamma}\big),\nonumber \\
\Delta= & \frac{1}{2}\left(\Gamma^{M}\partial_{M}\phi+\frac{1}{2\cdot3!}H_{MNK}\Gamma^{MNK}\bar{\gamma}\right)+\frac{1}{2}e^{\phi}\left(F_{M}\Gamma^{M}\bar{\gamma}+\frac{1}{2\cdot3!}F_{KLN}\Gamma^{KLN}\right),\nonumber \\
\Gamma_{\mathrm{D}_{p}}= & \frac{1}{\sqrt{-\left(g+\mathcal{F}\right)}}\sum_{q}\frac{\epsilon^{\alpha_{1}...\alpha_{2q}\beta_{1}...\beta_{p-2q+1}}}{q!2^{q}\left(p-2q+1\right)!}\mathcal{F}_{\alpha_{1}\alpha_{2}}...\mathcal{F}_{\alpha_{2q-1}\alpha_{2q}}\Gamma_{\beta_{1}...\beta_{p-2q+1}}\bar{\gamma}^{\frac{p-2q+1}{2}},\nonumber \\
\mathrm{L}_{\mathrm{D}_{p}} & =\sum_{q}\frac{\epsilon^{\alpha_{1}...\alpha_{2q}\beta_{1}...\beta_{p-2q+1}}}{q!2^{q}\left(p-2q+1\right)!}\frac{\left(-i\sigma_{2}\right)\left(\bar{\gamma}\right)^{\frac{p-2q+1}{2}}}{\sqrt{-\left(g+\mathcal{F}\right)}}\mathcal{F}_{\alpha_{1}\alpha_{2}}...\mathcal{F}_{\alpha_{2q-1}\alpha_{2q}}\Gamma_{\beta_{1}...\beta_{p-2q+1}}^{\ \ \ \ \ \ \ \ \ \ \ \ \ \lambda}\hat{D}_{\lambda}.\tag{B-6}\label{eq:B-6}
\end{align}
The indices denoted by the capital letters $K,L,M,N...$ and the lowercase
letters $a,b,...$ run over the ten-dimensional spacetime and its
tangent space respectively. The Greek alphabet $\alpha,\beta,\lambda$
denote the indices running over the worldvolume of the $\mathrm{D}_{p}$-brane.
The metric has to be written in terms of vielbein $e_{M}^{a}$ as
$g_{MN}=e_{M}^{a}\eta_{ab}e_{N}^{b}$ for spinors inducing the gamma
matrices defined as,

\begin{equation}
\left\{ \gamma^{a},\gamma^{b}\right\} =2\eta^{ab},\left\{ \Gamma^{M},\Gamma^{N}\right\} =2g^{MN},\tag{B-7}
\end{equation}
with $e_{M}^{a}\Gamma^{M}=\gamma^{a}$. $\omega_{\alpha ab}$ is the
spin connection presented in $\nabla_{\alpha}=\partial_{\alpha}+\frac{1}{4}\omega_{\alpha ab}\gamma^{ab}$
as the covariant derivative for fermion. The gamma matrix with multiple
indices can be defined by ranking alternate anti-symmetrically or
symmetrically the indices e.g. 
\begin{equation}
\gamma^{ab}=\frac{1}{2}\left[\gamma^{a},\gamma^{b}\right],\gamma^{abc}=\frac{1}{2}\left\{ \gamma^{a},\gamma^{bc}\right\} ,\gamma^{abcd}=\frac{1}{2}\left[\gamma^{a},\gamma^{bcd}\right]...\tag{B-8}
\end{equation}
$\Gamma^{MNK...}$ is defined as $e_{a}^{M}e_{b}^{N}e_{c}^{K}\gamma^{abc...}$
and $\bar{\gamma}=\gamma^{01...9}$. $\sigma_{2}$ refers to the associated
Pauli matrix. The worldvolume field $\mathcal{F}$ is defined as $\mathcal{F}_{\alpha\beta}=B_{\alpha\beta}+\left(2\pi\alpha^{\prime}\right)f_{\alpha\beta}$
where $B$ is the NS-NS 2-form in type II string theory with $H=dB$
and $f$ is the Yang-Mills field strength. The various $F_{M},F_{MN},F_{KLM}...$
are the associated field strengths of the massless Romand-Romand fields
presented in IIA or IIB string theory.

\subsection*{Appendix C: The actions involving the worldvolume fermions and bosons}

In this appendix, we collect the full formulas which should be presented
in the actions in (\ref{eq:4.8}). Since these formulas are very lengthy,
we define several auxiliary operators and functions to outline the
relations of these formulas. First, the derivative operators are defined
as,

\begin{align}
\hat{D}_{\mu} & =\partial_{\mu}+\Omega_{\mu}+\Delta_{\mu},\tag{C-1}\label{eq:C-1}
\end{align}
where the index $\mu=0,1,2,4$ ($x^{0}=t,x^{1}=x,,x^{2}=y,x^{4}=\zeta$)
and

\begin{equation}
\Omega_{\mu}=\begin{cases}
A_{X}\gamma_{\mu}\gamma^{4}+B_{X}\gamma^{2}\gamma_{\mu}+C_{X}\gamma_{\mu}, & \mu=0,1,2\\
0. & \mu=4
\end{cases}\tag{C-2}
\end{equation}
\[
\]

\[
\Delta_{\mu}=\begin{cases}
D_{X}\gamma_{\mu}\gamma^{4}+E_{X}\gamma_{\mu}, & \mu=0,1\\
D_{Y}\gamma_{2}\gamma^{4}+E_{Y}\gamma_{2}, & \mu=2\\
P\gamma^{2}\gamma^{4}+Q\gamma^{4}, & \mu=4
\end{cases}\tag{C-3}
\]
with the presented functions
\begin{align}
A_{X} & =-\frac{\zeta u^{\prime}}{2u^{2}},B_{X}=-\frac{1}{8}a,C_{X}=\frac{u}{16L^{2}},D_{X}=\frac{\zeta u^{\prime}\phi}{8u^{2}},E_{X}=-\frac{u\phi}{64L^{2}},\nonumber \\
D_{Y} & =-\left(\frac{\zeta\phi^{\prime}}{4u}-\frac{3}{8}\phi\frac{\zeta u^{\prime}}{u^{2}}\right),E_{Y}=E_{X},P=-\frac{au}{8}\frac{e^{\frac{7}{4}\phi}}{\zeta},Q=\frac{1}{16}\frac{\zeta}{L^{2}}e^{-\frac{1}{2}\phi}.\tag{C-4}
\end{align}
Then we denote the reduced fermion presented in (\ref{eq:3.37}) as

\begin{equation}
\psi=\left(\begin{array}{c}
\psi_{+}\\
\psi_{-}
\end{array}\right)=\left(\begin{array}{c}
\sum_{n}\psi_{+}^{\left(n\right)}f_{+}^{\left(n\right)}\\
\sum_{m}\psi_{-}^{\left(m\right)}f_{-}^{\left(m\right)}
\end{array}\right).\tag{C-5}\label{eq:C-5}
\end{equation}
In the following terms, all the indices of Greek alphabet run over
0,1 and ``$\prime$'' refers to the derivative with respect to $x^{4}=\zeta$.
Using (\ref{eq:C-1}) - (\ref{eq:C-5}), the full actions presented
in (\ref{eq:4.8}) are collected as

\begin{align}
S_{B,\psi}^{\left(1\right)}= & i\int d\zeta Xf_{\alpha\beta}\eta^{\beta\lambda}\bar{\psi}(1-\gamma^{3})\gamma^{3}\gamma^{\alpha}\hat{D}_{\lambda}\psi\nonumber \\
= & i\int d\zeta Xf_{01}\bigg[-i\psi_{-}^{\dagger}\tau^{2}\partial_{0}\psi_{+}-i\psi_{+}^{\dagger}\tau^{2}\partial_{0}\psi_{-}+i\psi_{+}^{\dagger}\tau^{1}\partial_{0}\psi_{+}\nonumber \\
 & -i\psi_{-}^{\dagger}\tau^{1}\partial_{0}\psi_{-}-\psi_{-}^{\dagger}\tau^{3}\partial_{1}\psi_{+}+\psi_{+}^{\dagger}\tau^{3}\partial_{1}\psi_{-}+i\psi_{+}^{\dagger}\partial_{1}\psi_{+}+i\psi_{-}^{\dagger}\partial_{1}\psi_{-}\nonumber \\
 & +2iB_{X}\left(\psi_{-}^{\dagger}\psi_{+}-\psi_{+}^{\dagger}\psi_{-}\right)+2\left(A_{X}+D_{X}-C_{X}-E_{X}\right)\psi_{-}^{\dagger}\tau^{2}\psi_{-}\nonumber \\
 & -2\left(A_{X}+D_{X}+C_{X}+E_{X}\right)\psi_{+}^{\dagger}\tau^{2}\psi_{+}+2B_{X}\left(\psi_{+}^{\dagger}\tau^{3}\psi_{+}+\psi_{-}^{\dagger}\tau^{3}\psi_{-}\right)\nonumber \\
 & -2\left(A_{X}+D_{X}-C_{X}-E_{X}\right)\psi_{+}^{\dagger}\tau^{1}\psi_{-}-2\left(A_{X}+D_{X}+C_{X}+E_{X}\right)\psi_{-}^{\dagger}\tau^{1}\psi_{+}\bigg],\tag{C-6}
\end{align}
where

\begin{equation}
X=\frac{T_{7}}{2}\left(2\pi\alpha^{\prime}\right)^{3}\sqrt{-g}e^{-\phi}\mathcal{A}^{-1}\left(\frac{u}{L}\right)^{3}=\frac{e^{\frac{\phi}{4}}\sqrt{\pi}u^{3}\zeta^{4}}{\sqrt{\lambda}\rho^{5}}.\tag{C-7}
\end{equation}
and
\begin{align}
S_{B,\psi}^{\left(2\right)}= & i\int d\zeta X\bar{\psi}(1-\gamma^{3})\gamma^{3}\gamma^{\alpha}f_{\alpha2}e^{\phi}\hat{D}_{2}\psi\nonumber \\
= & i\int d\zeta Xe^{\phi}\bigl(f_{02}\mathcal{S}+f_{12}\mathcal{R}\bigr),\tag{C-8}
\end{align}
where

\begin{align}
\mathcal{S}= & i\psi_{+}^{\dagger}\partial_{2}\psi_{+}+i\psi_{-}^{\dagger}\partial_{2}\psi_{-}-\psi_{-}^{\dagger}\tau^{3}\partial_{2}\psi_{+}+\psi_{+}^{\dagger}\tau^{3}\partial_{2}\psi_{-}\nonumber \\
 & +iB_{X}\left(\psi_{+}^{\dagger}\psi_{+}+\psi_{-}^{\dagger}\psi_{-}\right)+B_{X}\left(-\psi_{-}^{\dagger}\tau^{3}\psi_{+}+\psi_{+}^{\dagger}\tau^{3}\psi_{-}\right)\nonumber \\
 & +\left(A_{X}+D_{Y}+C_{X}+E_{X}\right)\psi_{-}^{\dagger}\tau^{1}\psi_{-}+\left(C_{X}+E_{Y}-A_{X}-D_{Y}\right)\psi_{+}^{\dagger}\tau^{1}\psi_{+}\nonumber \\
 & -\left(C_{X}+E_{Y}-A_{X}-D_{Y}\right)\psi_{-}^{\dagger}\tau^{2}\psi_{+}+\left(A_{X}+D_{Y}+C_{X}+E_{Y}\right)\psi_{+}^{\dagger}\tau^{2}\psi_{-},\nonumber \\
\mathcal{R}= & -i\psi_{-}^{\dagger}\tau^{2}\partial_{2}\psi_{+}-i\psi_{-}^{\dagger}\tau^{1}\partial_{2}\psi_{-}+i\psi_{+}^{\dagger}\tau^{1}\partial_{2}\psi_{+}-i\psi_{+}^{\dagger}\tau^{2}\partial_{2}\psi_{-}\nonumber \\
 & -iB_{X}\bigl(\psi_{-}^{\dagger}\tau^{2}\psi_{+}+\psi_{-}^{\dagger}\tau^{1}\psi_{-}-\psi_{+}^{\dagger}\tau^{1}\psi_{+}+\psi_{+}^{\dagger}\tau^{2}\psi_{-}\bigr)\nonumber \\
 & -\left(A_{X}+D_{Y}+C_{X}+E_{Y}\right)\psi_{-}^{\dagger}\psi_{-}+\left(C_{X}+E_{Y}-A_{X}-D_{Y}\right)\psi_{+}^{\dagger}\psi_{+}\nonumber \\
 & +i\left(C_{X}+E_{Y}-A_{X}-D_{Y}\right)\psi_{-}^{\dagger}\tau^{3}\psi_{+}+i\left(A_{X}+D_{Y}+C_{X}+E_{Y}\right)\psi_{+}^{\dagger}\tau^{3}\psi_{-}.\tag{C-9}
\end{align}
And

\begin{align}
S_{B,\psi}^{\left(3\right)}= & i\int d^{3}xd\zeta X\bar{\psi}(1-\gamma^{3})\gamma^{3}e^{\phi/2}\gamma^{2}f_{2\alpha}\eta^{\alpha\beta}\hat{D}_{\beta}\psi\nonumber \\
= & i\int d^{3}xd\zeta Xe^{\phi/2}\bigl(f_{20}\mathcal{P}+f_{21}\mathcal{Q}\bigr),\tag{C-10}
\end{align}
where

\begin{align}
\mathcal{P}= & -i\bigg[\psi_{-}^{\dagger}\tau^{1}\partial_{0}\psi_{+}+\psi_{+}^{\dagger}\tau^{2}\partial_{0}\psi_{+}-\psi_{-}^{\dagger}\tau^{2}\partial_{0}\psi_{-}+\psi_{+}^{\dagger}\tau^{1}\partial_{0}\psi_{-}\nonumber \\
 & -iB_{X}\psi_{-}^{\dagger}\tau^{3}\psi_{+}-B_{X}\psi_{+}^{\dagger}\psi_{+}-B_{X}\psi_{-}^{\dagger}\psi_{-}+iB_{X}\psi_{+}^{\dagger}\tau^{3}\psi_{-}\nonumber \\
 & +i\left(A_{X}+D_{X}-C_{X}-D_{X}\right)\psi_{-}^{\dagger}\tau^{1}\psi_{-}+i\left(A_{X}+D_{X}-C_{X}-D_{X}\right)\psi_{+}^{\dagger}\tau^{2}\psi_{-}\nonumber \\
 & +i\left(A_{X}+D_{X}+C_{X}+D_{X}\right)\psi_{-}^{\dagger}\tau^{2}\psi_{+}-i\left(A_{X}+D_{X}+C_{X}+D_{X}\right)\psi_{+}^{\dagger}\tau^{1}\psi_{+}\bigg],\nonumber \\
\mathcal{Q}= & i\bigg[\psi_{-}^{\dagger}\tau^{1}\partial_{1}\psi_{+}+\psi_{+}^{\dagger}\tau^{2}\partial_{1}\psi_{+}-\psi_{-}^{\dagger}\tau^{2}\partial_{1}\psi_{-}+\psi_{+}^{\dagger}\tau^{1}\partial_{1}\psi_{-}\nonumber \\
 & -B_{X}\psi_{-}^{\dagger}\tau^{2}\psi_{+}+B_{X}\psi_{+}^{\dagger}\tau^{1}\psi_{+}-B_{X}\psi_{-}^{\dagger}\tau^{1}\psi_{-}-B_{X}\psi_{+}^{\dagger}\tau^{2}\psi_{-}\nonumber \\
 & +i\left(A_{X}+D_{X}-C_{X}-D_{X}\right)\psi_{-}^{\dagger}\psi_{-}+\left(A_{X}+D_{X}-C_{X}-D_{X}\right)\psi_{+}^{\dagger}\tau^{3}\psi_{-}\nonumber \\
 & -\left(A_{X}+D_{X}+C_{X}+D_{X}\right)\psi_{-}^{\dagger}\tau^{3}\psi_{+}+i\left(A_{X}+D_{X}+C_{X}+D_{X}\right)\psi_{+}^{\dagger}\psi_{+}\bigg],\tag{C-11}
\end{align}
and

\begin{align}
S_{B,\psi,\pi}^{\left(1\right)}= & i\int d^{3}xd\zeta X\frac{\zeta}{u}e^{-\frac{\phi}{4}}\bar{\psi}(1-\gamma^{3})\gamma^{3}\gamma^{4}f_{4\alpha}\eta^{\alpha\beta}\hat{D}_{\beta}\psi\nonumber \\
= & i\int d^{3}xd\zeta X\frac{\zeta}{u}e^{-\frac{\phi}{4}}f_{4\alpha}\eta^{\alpha\beta}\mathcal{U}_{\beta},\tag{C-12}
\end{align}
where

\begin{align}
\mathcal{U}_{0}= & \psi_{-}^{\dagger}\partial_{0}\psi_{+}-i\psi_{+}^{\dagger}\tau^{3}\partial_{0}\psi_{+}-i\psi_{-}^{\dagger}\tau^{3}\partial_{0}\psi_{-}-\psi_{+}^{\dagger}\partial_{0}\psi_{-}\nonumber \\
 & -B_{X}\left(\psi_{-}^{\dagger}\tau^{2}\psi_{+}-\psi_{+}^{\dagger}\tau^{1}\psi_{+}+\psi_{-}^{\dagger}\tau^{1}\psi_{-}+\psi_{+}^{\dagger}\tau^{2}\psi_{-}\right)\nonumber \\
 & +i\left(A_{X}+D_{X}-C_{X}-E_{X}\right)\left(\psi_{-}^{\dagger}\psi_{-}+\psi_{+}^{\dagger}\tau^{3}\psi_{-}\right)\nonumber \\
 & -\left(A_{X}+D_{X}+C_{X}+E_{X}\right)\left(\psi_{-}^{\dagger}\tau^{3}\psi_{+}-i\psi_{+}^{\dagger}\psi_{+}\right),\nonumber \\
\mathcal{U}_{1}= & \psi_{-}^{\dagger}\partial_{1}\psi_{+}-i\psi_{+}^{\dagger}\tau^{3}\partial_{1}\psi_{+}-i\psi_{-}^{\dagger}\tau^{3}\partial_{1}\psi_{-}-\psi_{+}^{\dagger}\partial_{1}\psi_{-}\nonumber \\
 & -B_{X}\left(i\psi_{-}^{\dagger}\tau^{3}\psi_{+}+\psi_{+}^{\dagger}\psi_{+}+\psi_{-}^{\dagger}\psi_{-}-i\psi_{+}^{\dagger}\tau^{3}\psi_{-}\right)\nonumber \\
 & +i\left(A_{X}+D_{X}-C_{X}-E_{X}\right)\left(\psi_{-}^{\dagger}\tau^{1}\psi_{-}+\psi_{+}^{\dagger}\tau^{2}\psi_{-}\right)\nonumber \\
 & +i\left(A_{X}+D_{X}+C_{X}+E_{X}\right)\left(\psi_{-}^{\dagger}\tau^{2}\psi_{+}-\psi_{+}^{\dagger}\tau^{1}\psi_{+}\right).\tag{C-13}
\end{align}
And

\begin{align}
S_{B,\psi,\pi}^{\left(2\right)}= & i\int d^{3}xd\zeta X\sqrt{A}\bar{\psi}(1-\gamma^{3})\gamma^{3}\frac{\zeta^{2}}{u^{2}}e^{-\phi/2}\gamma^{\alpha}f_{\alpha4}\hat{D}_{4}\left(A^{-1/2}\psi\right)\nonumber \\
= & i\int d^{3}xd\zeta X\frac{\zeta^{2}}{u^{2}}e^{-\phi/2}\bigg\{ f_{04}\bigg[-i\psi_{-}^{\dagger}\partial_{4}\psi_{-}-i\psi_{+}^{\dagger}\partial_{4}\psi_{+}-\psi_{-}^{\dagger}\sigma^{3}\partial_{4}\psi_{+}-\psi_{+}^{\dagger}\sigma^{3}\partial_{4}\psi_{-}\nonumber \\
 & +Q\left(\psi_{+}^{\dagger}\sigma^{3}\psi_{-}-\psi_{-}^{\dagger}\sigma^{3}\psi_{+}+i\psi_{-}^{\dagger}\psi_{-}-i\psi_{+}^{\dagger}\psi_{+}\right)\nonumber \\
 & +P\left(\psi_{-}^{\dagger}\sigma^{2}\psi_{+}-\psi_{+}^{\dagger}\sigma^{2}\psi_{-}+\psi_{-}^{\dagger}\sigma^{1}\psi_{-}-\psi_{+}^{\dagger}\sigma^{1}\psi_{+}\right)\nonumber \\
 & +\frac{1}{2}\frac{A'}{A}\left(i\psi_{-}^{\dagger}\psi_{-}+i\psi_{+}^{\dagger}\psi_{+}+\psi_{-}^{\dagger}\sigma^{3}\psi_{+}+\psi_{+}^{\dagger}\sigma^{3}\psi_{-}\right)\bigg]\nonumber \\
 & +f_{14}\bigg[i\left(\psi_{-}^{\dagger}\sigma^{2}\partial_{4}\psi_{+}+\psi_{+}^{\dagger}\sigma^{2}\partial_{4}\psi_{-}-\psi_{-}^{\dagger}\sigma^{1}\partial_{4}\psi_{-}-\psi_{+}^{\dagger}\sigma^{1}\partial_{4}\psi_{+}\right)\nonumber \\
 & +iQ\left(\psi_{-}^{\dagger}\sigma^{2}\psi_{+}-\psi_{+}^{\dagger}\sigma^{2}\psi_{-}+\psi_{-}^{\dagger}\sigma^{1}\psi_{-}-\psi_{+}^{\dagger}\sigma^{1}\psi_{+}\right)\nonumber \\
 & +P\left(i\psi_{-}^{\dagger}\sigma^{3}\psi_{+}-i\psi_{+}^{\dagger}\sigma^{3}\psi_{-}+\psi_{-}^{\dagger}\psi_{-}-\psi_{+}^{\dagger}\psi_{+}\right)\nonumber \\
 & -\frac{1}{2}\frac{A'}{A}i\left(\psi_{-}^{\dagger}\sigma^{2}\psi_{+}+\psi_{+}^{\dagger}\sigma^{2}\psi_{-}-\psi_{-}^{\dagger}\sigma^{1}\psi_{-}-\psi_{+}^{\dagger}\sigma^{1}\psi_{+}\right)\bigg]\bigg\},\tag{C-14}
\end{align}
and

\begin{align}
S_{B,\psi,\pi}^{\left(3\right)}= & i\int d^{3}xd\zeta X\sqrt{A}\frac{\zeta^{2}}{u^{2}}\bar{\psi}\left(1-\gamma^{3}\right)\gamma^{3}\gamma^{2}f_{24}\hat{D}_{4}\left(A^{-1/2}\psi\right)\nonumber \\
= & i\int d^{3}xd\zeta X\frac{\zeta^{2}}{u^{2}}f_{24}\Bigg\{ i\left(\psi_{-}^{\dagger}\tau^{1}\partial_{4}\psi_{+}+\psi_{+}^{\dagger}\tau^{1}\partial_{4}\psi_{-}+\psi_{+}^{\dagger}\tau^{2}\partial_{4}\psi_{+}-\psi_{-}^{\dagger}\tau^{2}\partial_{4}\psi_{-}\right)\nonumber \\
 & -i\frac{A^{\prime}}{2A}\left(\psi_{-}^{\dagger}\tau^{1}\psi_{+}+\psi_{+}^{\dagger}\tau^{1}\psi_{-}+\psi_{+}^{\dagger}\tau^{2}\psi_{+}-\psi_{-}^{\dagger}\tau^{2}\psi_{-}\right)\nonumber \\
 & -P\left(i\psi_{-}^{\dagger}\tau^{3}\psi_{-}+i\psi_{+}^{\dagger}\tau^{3}\psi_{+}-\psi_{-}^{\dagger}\psi_{+}+\psi_{+}^{\dagger}\psi_{-}\right)\nonumber \\
 & +iQ\left(\psi_{-}^{\dagger}\tau^{1}\psi_{+}-\psi_{+}^{\dagger}\tau^{1}\psi_{-}+\psi_{+}^{\dagger}\tau^{2}\psi_{+}+\psi_{-}^{\dagger}\tau^{2}\psi_{-}\right)\Bigg\},\tag{C-15}
\end{align}
and

\begin{align}
S_{B,\psi,\pi}^{\left(4\right)}= & i\int d^{3}xd\zeta X\frac{\zeta}{u}e^{\frac{3\phi}{4}}\bar{\psi}\left(1-\gamma^{3}\right)\gamma^{3}\gamma^{4}f_{42}\hat{D}_{2}\psi\nonumber \\
= & i\int d^{3}xd\zeta X\frac{\zeta}{u}e^{3\phi/4}f_{42}\bigg[\psi_{-}^{\dagger}\partial_{2}\psi_{+}-\psi_{+}^{\dagger}\partial_{2}\psi_{-}-i\psi_{-}^{\dagger}\tau^{3}\partial_{2}\psi_{-}-i\psi_{+}^{\dagger}\tau^{3}\partial_{2}\psi_{+}\nonumber \\
 & -i\left(A_{X}+D_{Y}+C_{X}+E_{Y}\right)\left(\psi_{-}^{\dagger}\tau^{1}\psi_{+}+\psi_{+}^{\dagger}\tau^{2}\psi_{+}\right)\nonumber \\
 & -i\left(A_{X}+D_{Y}-C_{X}-E_{Y}\right)\left(\psi_{+}^{\dagger}\tau^{1}\psi_{-}-\psi_{-}^{\dagger}\tau^{2}\psi_{-}\right)\nonumber \\
 & +B_{X}\left(\psi_{-}^{\dagger}\psi_{+}-\psi_{+}^{\dagger}\psi_{-}-i\psi_{-}^{\dagger}\tau^{3}\psi_{-}-i\psi_{+}^{\dagger}\tau^{3}\psi_{+}\right)\bigg].\tag{C-16}
\end{align}

\bibliographystyle{utphys}
\bibliography{HQCD3_anisotropy}

@article{Taubes:1999bv,
    author = "Taubes, G.",
    title = "{String theorists find a rosetta stone}",
    doi = "10.1126/science.285.5427.512",
    journal = "Science",
    volume = "285",
    pages = "512--517",
    year = "1999"
}

@article{Aharony:1999ti,
    author = "Aharony, Ofer and Gubser, Steven S. and Maldacena, Juan Martin and Ooguri, Hirosi and Oz, Yaron",
    title = "{Large N field theories, string theory and gravity}",
    eprint = "hep-th/9905111",
    archivePrefix = "arXiv",
    reportNumber = "CERN-TH-99-122, HUTP-99-A027, LBNL-43113, RU-99-18, UCB-PTH-99-16, LBL-43113",
    doi = "10.1016/S0370-1573(99)00083-6",
    journal = "Phys. Rept.",
    volume = "323",
    pages = "183--386",
    year = "2000"
}

@article{Witten:1998qj,
    author = "Witten, Edward",
    title = "{Anti de Sitter space and holography}",
    eprint = "hep-th/9802150",
    archivePrefix = "arXiv",
    reportNumber = "IASSNS-HEP-98-15",
    doi = "10.4310/ATMP.1998.v2.n2.a2",
    journal = "Adv. Theor. Math. Phys.",
    volume = "2",
    pages = "253--291",
    year = "1998"
}

@article{Maldacena:1997re,
    author = "Maldacena, Juan Martin",
    title = "{The Large $N$ limit of superconformal field theories and supergravity}",
    eprint = "hep-th/9711200",
    archivePrefix = "arXiv",
    reportNumber = "HUTP-97-A097, HUTP-98-A097",
    doi = "10.4310/ATMP.1998.v2.n2.a1",
    journal = "Adv. Theor. Math. Phys.",
    volume = "2",
    pages = "231--252",
    year = "1998"
}

@article{Shuryak:2003xe,
    author = "Shuryak, Edward",
    editor = "Faessler, A.",
    title = "{Why does the quark gluon plasma at RHIC behave as a nearly ideal fluid?}",
    eprint = "hep-ph/0312227",
    archivePrefix = "arXiv",
    doi = "10.1016/j.ppnp.2004.02.025",
    journal = "Prog. Part. Nucl. Phys.",
    volume = "53",
    pages = "273--303",
    year = "2004"
}

@article{Shuryak:2004cy,
    author = "Shuryak, Edward V.",
    editor = "Rischke, D. and Levin, G.",
    title = "{What RHIC experiments and theory tell us about properties of quark-gluon plasma?}",
    eprint = "hep-ph/0405066",
    archivePrefix = "arXiv",
    doi = "10.1016/j.nuclphysa.2004.10.022",
    journal = "Nucl. Phys. A",
    volume = "750",
    pages = "64--83",
    year = "2005"
}

@article{Florkowski:2010cf,
    author = "Florkowski, Wojciech and Ryblewski, Radoslaw",
    title = "{Highly-anisotropic and strongly-dissipative hydrodynamics for early stages of relativistic heavy-ion collisions}",
    eprint = "1007.0130",
    archivePrefix = "arXiv",
    primaryClass = "nucl-th",
    doi = "10.1103/PhysRevC.83.034907",
    journal = "Phys. Rev. C",
    volume = "83",
    pages = "034907",
    year = "2011"
}

@article{Ryblewski:2010bs,
    author = "Ryblewski, Radoslaw and Florkowski, Wojciech",
    title = "{Non-boost-invariant motion of dissipative and highly anisotropic fluid}",
    eprint = "1007.4662",
    archivePrefix = "arXiv",
    primaryClass = "nucl-th",
    doi = "10.1088/0954-3899/38/1/015104",
    journal = "J. Phys. G",
    volume = "38",
    pages = "015104",
    year = "2011"
}

@article{Martinez:2010sc,
    author = "Martinez, Mauricio and Strickland, Michael",
    title = "{Dissipative Dynamics of Highly Anisotropic Systems}",
    eprint = "1007.0889",
    archivePrefix = "arXiv",
    primaryClass = "nucl-th",
    doi = "10.1016/j.nuclphysa.2010.08.011",
    journal = "Nucl. Phys. A",
    volume = "848",
    pages = "183--197",
    year = "2010"
}

@article{Martinez:2010sd,
    author = "Martinez, Mauricio and Strickland, Michael",
    title = "{Non-boost-invariant anisotropic dynamics}",
    eprint = "1011.3056",
    archivePrefix = "arXiv",
    primaryClass = "nucl-th",
    doi = "10.1016/j.nuclphysa.2011.02.003",
    journal = "Nucl. Phys. A",
    volume = "856",
    pages = "68--87",
    year = "2011"
}

@article{Giataganas:2017koz,
    author = {Giataganas, Dimitrios and G{\"u}rsoy, Umut and Pedraza, Juan F.},
    title = "{Strongly-coupled anisotropic gauge theories and holography}",
    eprint = "1708.05691",
    archivePrefix = "arXiv",
    primaryClass = "hep-th",
    reportNumber = "NCTS-TH/1712, NCTS-TH-1712",
    doi = "10.1103/PhysRevLett.121.121601",
    journal = "Phys. Rev. Lett.",
    volume = "121",
    number = "12",
    pages = "121601",
    year = "2018"
}

@article{Banks:2016fab,
    author = "Banks, Elliot",
    title = "{Phase transitions of an anisotropic N=4 super Yang-Mills plasma via holography}",
    eprint = "1604.03552",
    archivePrefix = "arXiv",
    primaryClass = "hep-th",
    doi = "10.1007/JHEP07(2016)085",
    journal = "JHEP",
    volume = "07",
    pages = "085",
    year = "2016"
}

@article{Avila:2016mno,
    author = "{\'A}vila, Daniel and Fern{\'a}ndez, Daniel and Pati{\~n}o, Leonardo and Trancanelli, Diego",
    title = "{Thermodynamics of anisotropic branes}",
    eprint = "1609.02167",
    archivePrefix = "arXiv",
    primaryClass = "hep-th",
    reportNumber = "MPP-2016-275",
    doi = "10.1007/JHEP11(2016)132",
    journal = "JHEP",
    volume = "11",
    pages = "132",
    year = "2016"
}

@article{Li:2022wwv,
    author = "Li, Si-wen and Luo, Sen-kai and Hu, Ya-qian",
    title = "{Holographic QCD$_{3}$ and Chern-Simons theory from anisotropic supergravity}",
    eprint = "2203.14489",
    archivePrefix = "arXiv",
    primaryClass = "hep-th",
    doi = "10.1007/JHEP06(2022)040",
    journal = "JHEP",
    volume = "06",
    pages = "040",
    year = "2022"
}

@article{Li:2025ahp,
    author = "Li, Si-wen and Zhang, Xiao-tong",
    title = "{Worldvolume fermion as baryon with homogeneous instantons in holographic QCD3}",
    eprint = "2503.24173",
    archivePrefix = "arXiv",
    primaryClass = "hep-th",
    doi = "10.1103/d1p6-4yl9",
    journal = "Phys. Rev. D",
    volume = "112",
    number = "2",
    pages = "026001",
    year = "2025"
}

@article{Li:2021vve,
    author = "Li, Si-wen and Luo, Sen-kai and Tan, Mu-zhi",
    title = "{Three-dimensional Yang-Mills-Chern-Simons theory from a D3-brane background with D-instantons}",
    eprint = "2106.04038",
    archivePrefix = "arXiv",
    primaryClass = "hep-th",
    doi = "10.1103/PhysRevD.104.066008",
    journal = "Phys. Rev. D",
    volume = "104",
    number = "6",
    pages = "066008",
    year = "2021"
}

@article{Argurio:2020her,
    author = "Argurio, Riccardo and Armoni, Adi and Bertolini, Matteo and Mignosa, Francesco and Niro, Pierluigi",
    title = "{Vacuum structure of large $N$ $QCD_{3}$ from holography}",
    eprint = "2006.01755",
    archivePrefix = "arXiv",
    primaryClass = "hep-th",
    doi = "10.1007/JHEP07(2020)134",
    journal = "JHEP",
    volume = "07",
    pages = "134",
    year = "2020"
}

@article{Hong:2010sb,
    author = "Hong, Deog Ki and Yee, Ho-Ung",
    title = "{Holographic aspects of three dimensional QCD from string theory}",
    eprint = "1003.1306",
    archivePrefix = "arXiv",
    primaryClass = "hep-th",
    reportNumber = "PNUTP-10-A01, IC-2010-007",
    doi = "10.1007/JHEP05(2010)036",
    journal = "JHEP",
    volume = "05",
    pages = "036",
    year = "2010",
    note = "[Erratum: JHEP 08, 120 (2010)]"
}

@article{Mateos:2011tv,
    author = "Mateos, David and Trancanelli, Diego",
    title = "{Thermodynamics and Instabilities of a Strongly Coupled Anisotropic Plasma}",
    eprint = "1106.1637",
    archivePrefix = "arXiv",
    primaryClass = "hep-th",
    reportNumber = "ICCUB-11-145, MAD-TH-11-03",
    doi = "10.1007/JHEP07(2011)054",
    journal = "JHEP",
    volume = "07",
    pages = "054",
    year = "2011"
}

@article{Rebhan:2011vd,
    author = "Rebhan, Anton and Steineder, Dominik",
    title = "{Violation of the Holographic Viscosity Bound in a Strongly Coupled Anisotropic Plasma}",
    eprint = "1110.6825",
    archivePrefix = "arXiv",
    primaryClass = "hep-th",
    doi = "10.1103/PhysRevLett.108.021601",
    journal = "Phys. Rev. Lett.",
    volume = "108",
    pages = "021601",
    year = "2012"
}

@article{Mateos:2011ix,
    author = "Mateos, David and Trancanelli, Diego",
    title = "{The anisotropic N=4 super Yang-Mills plasma and its instabilities}",
    eprint = "1105.3472",
    archivePrefix = "arXiv",
    primaryClass = "hep-th",
    reportNumber = "ICCUB-11-142, MAD-TH-10-06",
    doi = "10.1103/PhysRevLett.107.101601",
    journal = "Phys. Rev. Lett.",
    volume = "107",
    pages = "101601",
    year = "2011"
}

@article{DElia:2012pvq,
    author = "D'Elia, Massimo and Negro, Francesco",
    title = "{$\theta$ dependence of the deconfinement temperature in Yang-Mills theories}",
    eprint = "1205.0538",
    archivePrefix = "arXiv",
    primaryClass = "hep-lat",
    reportNumber = "IFUP-TH-2012-07",
    doi = "10.1103/PhysRevLett.109.072001",
    journal = "Phys. Rev. Lett.",
    volume = "109",
    pages = "072001",
    year = "2012"
}

@article{DElia:2013uaf,
    author = "D'Elia, Massimo and Negro, Francesco",
    title = "{Phase diagram of Yang-Mills theories in the presence of a $\theta$ term}",
    eprint = "1306.2919",
    archivePrefix = "arXiv",
    primaryClass = "hep-lat",
    reportNumber = "IFUP-TH-2013-13",
    doi = "10.1103/PhysRevD.88.034503",
    journal = "Phys. Rev. D",
    volume = "88",
    number = "3",
    pages = "034503",
    year = "2013"
}

@article{Schafer:1996wv,
    author = {Sch{\"a}fer, Thomas and Shuryak, Edward V.},
    title = "{Instantons in QCD}",
    eprint = "hep-ph/9610451",
    archivePrefix = "arXiv",
    reportNumber = "DOE-ER-40561-293, INT-96-00-150",
    doi = "10.1103/RevModPhys.70.323",
    journal = "Rev. Mod. Phys.",
    volume = "70",
    pages = "323--426",
    year = "1998"
}

@article{Gross:1980br,
    author = "Gross, David J. and Pisarski, Robert D. and Yaffe, Laurence G.",
    title = "{QCD and Instantons at Finite Temperature}",
    reportNumber = "PRINT-80-0538 (PRINCETON)",
    doi = "10.1103/RevModPhys.53.43",
    journal = "Rev. Mod. Phys.",
    volume = "53",
    pages = "43",
    year = "1981"
}

@inproceedings{Tong:2005un,
    author = "Tong, David",
    title = "{TASI lectures on solitons: Instantons, monopoles, vortices and kinks}",
    booktitle = "{Theoretical Advanced Study Institute in Elementary Particle Physics}: {Many Dimensions of String Theory}",
    eprint = "hep-th/0509216",
    archivePrefix = "arXiv",
    month = "6",
    year = "2005"
}

@article{Vicari:2008jw,
    author = "Vicari, Ettore and Panagopoulos, Haralambos",
    title = "{Theta dependence of SU(N) gauge theories in the presence of a topological term}",
    eprint = "0803.1593",
    archivePrefix = "arXiv",
    primaryClass = "hep-th",
    doi = "10.1016/j.physrep.2008.10.001",
    journal = "Phys. Rept.",
    volume = "470",
    pages = "93--150",
    year = "2009"
}

@article{HFLAV:2022esi,
    author = "Amhis, Yasmine Sara and others",
    collaboration = "HFLAV",
    title = "{Averages of b-hadron, c-hadron, and {\ensuremath{\tau}}-lepton properties as of 2021}",
    eprint = "2206.07501",
    archivePrefix = "arXiv",
    primaryClass = "hep-ex",
    doi = "10.1103/PhysRevD.107.052008",
    journal = "Phys. Rev. D",
    volume = "107",
    number = "5",
    pages = "052008",
    year = "2023"
}

@article{LHCb:2025ray,
    author = "Aaij, Roel and others",
    collaboration = "LHCb",
    title = "{Observation of charge{\textendash}parity symmetry breaking in baryon decays}",
    eprint = "2503.16954",
    archivePrefix = "arXiv",
    primaryClass = "hep-ex",
    reportNumber = "LHCb-PAPER-2024-054, CERN-EP-2025-031",
    doi = "10.1038/s41586-025-09119-3",
    journal = "Nature",
    volume = "643",
    number = "8074",
    pages = "1223--1228",
    year = "2025"
}

@article{Witten:1998zw,
    author = "Witten, Edward",
    editor = "Bergstrom, L. and Lindstrom, U.",
    title = "{Anti-de Sitter space, thermal phase transition, and confinement in gauge theories}",
    eprint = "hep-th/9803131",
    archivePrefix = "arXiv",
    reportNumber = "IASSNS-HEP-98-21",
    doi = "10.4310/ATMP.1998.v2.n3.a3",
    journal = "Adv. Theor. Math. Phys.",
    volume = "2",
    pages = "505--532",
    year = "1998"
}

@article{Becker:2007,
    author = "Becker, L. and Becker, M. and Schwarz, J.",
    editor = "Bergstrom, L. and Lindstrom, U.",
    title =  "{String Theory and M-Theory: A Modern Introduction}",
    journal = "Cambridge University Press",
    year = "2007"
}

@article{Witten:1998xy,
    author = "Witten, Edward",
    title = "{Baryons and branes in anti-de Sitter space}",
    eprint = "hep-th/9805112",
    archivePrefix = "arXiv",
    reportNumber = "IASSNS-HEP-98-42",
    doi = "10.1088/1126-6708/1998/07/006",
    journal = "JHEP",
    volume = "07",
    pages = "006",
    year = "1998"
}

@article{Marolf:2003ye,
    author = "Marolf, Donald and Martucci, Luca and Silva, Pedro J.",
    title = "{Fermions, T duality and effective actions for D-branes in bosonic backgrounds}",
    eprint = "hep-th/0303209",
    archivePrefix = "arXiv",
    reportNumber = "IFUM-751-FT",
    doi = "10.1088/1126-6708/2003/04/051",
    journal = "JHEP",
    volume = "04",
    pages = "051",
    year = "2003"
}

@article{Marolf:2003vf,
    author = "Marolf, Donald and Martucci, Luca and Silva, Pedro J.",
    title = "{Actions and Fermionic symmetries for D-branes in bosonic backgrounds}",
    eprint = "hep-th/0306066",
    archivePrefix = "arXiv",
    reportNumber = "IFUM-760-FT",
    doi = "10.1088/1126-6708/2003/07/019",
    journal = "JHEP",
    volume = "07",
    pages = "019",
    year = "2003"
}

@article{Li:2023wyb,
    author = "Li, Si-wen and Zhang, Yi-peng and Li, Hao-qian",
    title = "{Correlation function of flavored fermion in holographic QCD}",
    eprint = "2307.13357",
    archivePrefix = "arXiv",
    primaryClass = "hep-th",
    doi = "10.1103/PhysRevD.109.086020",
    journal = "Phys. Rev. D",
    volume = "109",
    number = "8",
    pages = "086020",
    year = "2024"
}

@article{Nakas:2020hyo,
    author = "Nakas, Theodoros and Rigatos, Konstantinos S.",
    title = "{Fermions and baryons as open-string states from brane junctions}",
    eprint = "2010.00025",
    archivePrefix = "arXiv",
    primaryClass = "hep-th",
    doi = "10.1007/JHEP12(2020)157",
    journal = "JHEP",
    volume = "12",
    pages = "157",
    year = "2020"
}

@article{physleta.2016,
title = {Effective Ginzburg-Landau free energy functional for multi-band isotropic superconductors},
journal = {Physics Letters A},
volume = {380},
number = {20},
pages = {1781-1787},
year = {2016},
issn = {0375-9601},
doi = {https://doi.org/10.1016/j.physleta.2016.03.023},
url = {https://www.sciencedirect.com/science/article/pii/S0375960116300020},
author = {Konstantin V. Grigorishin},
}

@article{PhysRevLett.83.5350,
  title = {Phenomenological Theory of Superconductivity and Magnetism in ${\mathrm{Ho}}_{1\ensuremath{-}\mathit{x}}{\mathrm{Dy}}_{\mathit{x}}{\mathrm{Ni}}_{2}{B}_{2}C$},
  author = {Doh, Hyeonjin and Sigrist, Manfred and Cho, B. K. and Lee, Sung-Ik},
  journal = {Phys. Rev. Lett.},
  volume = {83},
  issue = {25},
  pages = {5350--5353},
  numpages = {0},
  year = {1999},
  month = {Dec},
  publisher = {American Physical Society},
  doi = {10.1103/PhysRevLett.83.5350},
  url = {https://link.aps.org/doi/10.1103/PhysRevLett.83.5350}
}

@article{Li:2024jkd,
    author = "Li, Si-wen and Li, Hao-qian and Zhang, Yi-peng",
    title = "{Worldvolume fermions as baryons in holographic quantum chromodynamics with instantons}",
    eprint = "2402.01197",
    archivePrefix = "arXiv",
    primaryClass = "hep-th",
    doi = "10.1088/1572-9494/ad782f",
    journal = "Commun. Theor. Phys.",
    volume = "77",
    number = "1",
    pages = "015203",
    year = "2025"
}

@book{Casalderrey-Solana:2011dxg,
    author = "Casalderrey-Solana, Jorge and Liu, Hong and Mateos, David and Rajagopal, Krishna and Achim Wiedemann, Urs",
    title = "{Gauge/String Duality, Hot QCD and Heavy Ion Collisions}",
    eprint = "1101.0618",
    archivePrefix = "arXiv",
    primaryClass = "hep-th",
    reportNumber = "CERN-PH-TH-2010-316, MIT-CTP-4198, ICCUB-10-202",
    doi = "10.1017/9781009403504",
    isbn = "978-1-009-40350-4, 978-1-009-40349-8, 978-1-009-40352-8, 978-1-139-13674-7",
    publisher = "Cambridge University Press",
    year = "2014"
}

@article{Kruczenski:2003uq,
    author = "Kruczenski, Martin and Mateos, David and Myers, Robert C. and Winters, David J.",
    title = "{Towards a holographic dual of large N(c) QCD}",
    eprint = "hep-th/0311270",
    archivePrefix = "arXiv",
    doi = "10.1088/1126-6708/2004/05/041",
    journal = "JHEP",
    volume = "05",
    pages = "041",
    year = "2004"
}

@article{Sakai:2004cn,
    author = "Sakai, Tadakatsu and Sugimoto, Shigeki",
    title = "{Low energy hadron physics in holographic QCD}",
    eprint = "hep-th/0412141",
    archivePrefix = "arXiv",
    reportNumber = "IU-MSTP-63, YITP-04-70",
    doi = "10.1143/PTP.113.843",
    journal = "Prog. Theor. Phys.",
    volume = "113",
    pages = "843--882",
    year = "2005"
}

@inproceedings{Manohar:1998xv,
    author = "Manohar, Aneesh V.",
    title = "{Large N QCD}",
    booktitle = "{Les Houches Summer School in Theoretical Physics, Session 68: Probing the Standard Model of Particle Interactions}",
    eprint = "hep-ph/9802419",
    archivePrefix = "arXiv",
    reportNumber = "UCSD-PTH-98-06",
    pages = "1091--1169",
    month = "2",
    year = "1998"
}

@article{Dashen:1993jt,
    author = "Dashen, Roger F. and Jenkins, Elizabeth Ellen and Manohar, Aneesh V.",
    title = "{The 1/N(c) expansion for baryons}",
    eprint = "hep-ph/9310379",
    archivePrefix = "arXiv",
    reportNumber = "UCSD-PTH-93-21",
    doi = "10.1103/PhysRevD.51.2489",
    journal = "Phys. Rev. D",
    volume = "49",
    pages = "4713",
    year = "1994",
    note = "[Erratum: Phys.Rev.D 51, 2489 (1995)]"
}

@article{Camporesi:1995fb,
    author = "Camporesi, Roberto and Higuchi, Atsushi",
    title = "{On the Eigen functions of the Dirac operator on spheres and real hyperbolic spaces}",
    eprint = "gr-qc/9505009",
    archivePrefix = "arXiv",
    reportNumber = "BUTP-95-12",
    doi = "10.1016/0393-0440(95)00042-9",
    journal = "J. Geom. Phys.",
    volume = "20",
    pages = "1--18",
    year = "1996"
}

@article{Johnson:2002,
    author = "Clifford V Johnson",
    title = "{D-branes}",
    journal = "Cambridge University Press",
    year = "2002"
}

@article{Chernicoff:2012iq,
    author = "Chernicoff, Mariano and Fernandez, Daniel and Mateos, David and Trancanelli, Diego",
    title = "{Drag force in a strongly coupled anisotropic plasma}",
    eprint = "1202.3696",
    archivePrefix = "arXiv",
    primaryClass = "hep-th",
    reportNumber = "DAMTP-2012-12, ICCUB-12-088, MAD-TH-12-01",
    doi = "10.1007/JHEP08(2012)100",
    journal = "JHEP",
    volume = "08",
    pages = "100",
    year = "2012"
}

@article{Son:2002sd,
    author = "Son, Dam T. and Starinets, Andrei O.",
    title = "{Minkowski space correlators in AdS / CFT correspondence: Recipe and applications}",
    eprint = "hep-th/0205051",
    archivePrefix = "arXiv",
    reportNumber = "INT-PUB-02-34",
    doi = "10.1088/1126-6708/2002/09/042",
    journal = "JHEP",
    volume = "09",
    pages = "042",
    year = "2002"
}

@article{Li:2025zbj,
    author = "Li, Si-wen and Lin, Shu",
    title = "{Spin polarization of holographic baryon in strongly coupled fluid}",
    eprint = "2505.21883",
    archivePrefix = "arXiv",
    primaryClass = "hep-th",
    doi = "10.1103/jkv4-m723",
    journal = "Phys. Rev. D",
    volume = "112",
    number = "7",
    pages = "074009",
    year = "2025"
}

@article{Li:2024apc,
    author = "Li, Si-wen and Zhang, Yi-peng and Li, Hao-qian",
    title = "{Holographic spectroscopy of fermion with instantons}",
    eprint = "2406.11557",
    archivePrefix = "arXiv",
    primaryClass = "hep-th",
    doi = "10.1140/epjc/s10052-025-14556-5",
    journal = "Eur. Phys. J. C",
    volume = "85",
    number = "8",
    pages = "830",
    year = "2025"
}

\end{document}